\documentclass[11pt,a4paper]{article}
\usepackage[T1]{fontenc}
\usepackage[margin=25mm]{geometry}
\usepackage{amsmath,amssymb,amsthm,bm,mathtools,booktabs,array,graphicx,needspace}
\usepackage{caption}
\usepackage[hidelinks]{hyperref}

\newcommand{\Tr}{\operatorname{Tr}}
\newcommand{\tr}{\operatorname{tr}}
\newcommand{\rank}{\operatorname{rank}}
\newcommand{\ran}{\operatorname{ran}}
\newcommand{\diag}{\operatorname{diag}}
\newcommand{\R}{\mathbb R}
\newcommand{\C}{\mathbb C}
\newcommand{\HH}{\mathcal Q}

\newcommand{\ket}[1]{\lvert#1\rangle}
\newcommand{\bra}[1]{\langle#1\rvert}
\theoremstyle{definition}
\newtheorem{theorem}{Theorem}[section]
\newtheorem{proposition}[theorem]{Proposition}
\newtheorem{lemma}[theorem]{Lemma}

\renewcommand{\thetheorem}{\arabic{section}.\arabic{theorem}}
\numberwithin{equation}{section}
\renewcommand{\theequation}{\arabic{section}.\arabic{equation}}
\allowdisplaybreaks[1]
\hypersetup{pdftitle={Efficient numerical measurement design and certified risk bounds for multiparameter quantum estimation},pdfauthor={Koichi Yamagata and Wataru Ikeura}}
\begin{document}
\title{Efficient numerical measurement design and certified risk bounds for multiparameter quantum estimation}
\author{%
Koichi Yamagata\textsuperscript{1}\thanks{\href{mailto:yamagata@se.kanazawa-u.ac.jp}{yamagata@se.kanazawa-u.ac.jp}}
\quad Wataru Ikeura\textsuperscript{2}\\[0.7em]
\small \textsuperscript{1}Institute of Science and Engineering, Kanazawa University\\
\small Kanazawa, Ishikawa 920-1192, Japan\\[0.3em]
\small \textsuperscript{2}Graduate School of Natural Science and Technology, Kanazawa University\\
\small Kanazawa, Ishikawa 920-1192, Japan}
\date{November 11, 2026}
\maketitle
\begin{abstract}
Multiparameter quantum estimation requires efficient measurement design and reliable bounds on the attainable precision.
We present Q-StARC, a numerical framework that combines measurement optimization with certified lower bounds on the weighted variance of locally unbiased estimators.
The measurement search parameterizes rank-one measurements on a complex Stiefel manifold and combines projected ambient Adam updates, QR retraction, and Riemannian conjugate-gradient refinement.
Independent risk bounds are obtained from a denominator-based matrix sum-of-squares hierarchy.
When its support coefficients are optimized jointly, the hierarchy starts at the weighted Nagaoka--Hayashi (NH) bound and converges to the optimal locally unbiased risk, including for singular states.
Numerical applications certify that the NH bound is unattainable for several finite-copy qubit models and two full eight-parameter qutrit models.
For the qutrit examples, the certified relative gaps above the NH bound exceed $14\%$ and $11\%$, while the level-$m=1$ lower bounds agree with the best numerical measurement costs to relative differences of $4.5\times10^{-7}$ and $2.2\times10^{-6}$, respectively.
\end{abstract}
\section{Introduction}\label{sec:introduction}
The precision of a quantum measurement becomes a joint design problem when several unknown parameters must be estimated simultaneously.
Measurements that are optimal for individual parameters can be incompatible, and their allocation determines the tradeoff between estimation errors.
For a finite number of jointly measured copies, this tradeoff must be assessed within the available measurement space; an asymptotically attainable bound alone need not identify the precision attainable with those resources~\cite{MultiparameterReview,NH}.
The resulting optimization is relevant both to the design of quantum measurements and to the interpretation of precision bounds in multiparameter quantum metrology.

Locally unbiased estimation provides a precise setting for this problem.
For a fixed measurement with nonsingular classical Fisher information, a locally unbiased estimator attains the classical Cram\'er--Rao bound at the reference point.
The remaining task is therefore to minimize the weighted trace of the inverse Fisher information over all positive operator-valued measures (POVMs).
This formulation makes measurement design accessible to numerical optimization, including adaptive-estimation applications~\cite{KTF}, but a local search does not by itself certify global optimality.
An independent lower bound is needed to determine whether further improvement is possible and to distinguish a limitation of the search from a limitation imposed by quantum statistics.

Convex optimization has made several important quantum precision bounds computationally accessible.
The Holevo bound admits a semidefinite-programming formulation~\cite{AFD}, and the Nagaoka--Hayashi (NH) bound gives a computationally tractable relaxation of the locally unbiased measurement problem~\cite{NH}.
More generally, linear and conic formulations connect attainable precision to optimization over suitable cones~\cite{Hayashi,HO}.
In particular, Hayashi and Ouyang developed a framework for computing upper and lower bounds on the optimal risk and demonstrated numerical gaps from earlier computable bounds~\cite{HO}.
These results motivate concrete algorithms that combine efficient measurement searches with lower bounds whose validity can be checked independently of floating-point solver output.

In this work we develop Q-StARC, \emph{Quantum Estimator Design via Stiefel Adam--RCG Optimization with Certified Risk Bounds}.
Its measurement search uses a rank-one representation, which preserves the optimal value when sufficiently many outcomes are allowed~\cite{Yamagata}.
The measurement vectors form a complex Stiefel matrix, so positivity and completeness are incorporated directly into the parameterization.
Projected ambient Adam updates~\cite{Adam} and a thin QR retraction provide an initial search, followed by Riemannian conjugate-gradient (RCG) refinement~\cite{AMS}.
This representation avoids constructing dense effect matrices and reduces measurement storage from $O(sn^2)$ to $O(sn)$, where $n$ is the Hilbert-space dimension and $s$ is the number of outcomes.
The numerical comparison evaluates the two-stage update scheme as a whole; it does not isolate the contribution of each component to the observed performance.

The lower-bound construction is based on support functions of the attainable Fisher-information set.
We give a direct alternative derivation of the established quantum-estimation duality~\cite{Hayashi,HO} in a fixed-coefficient Fisher-support formulation, using supporting hyperplanes of a finite-dimensional concave value function.
The dual constraint is positivity of a quadratic matrix polynomial for every real value of its argument.
Multiplication by powers of $1+\|\xi\|^2$ leads to a hierarchy of matrix sum-of-squares (SOS) constraints, building on polynomial-optimization and matrix-SOS methods~\cite{PV,SH,Parrilo2003,DHL}.
We prove an exact reduction of the Gram representation, with different degree bounds in the support and kernel of the state, and construct strictly feasible reduced Gram matrices at every level.
The resulting convergence proof also applies to singular states.
When the support coefficients are optimized jointly with the Gram matrix, each level remains a single semidefinite program (SDP): the zeroth level equals the weighted NH bound, and the hierarchy converges monotonically to the optimal locally unbiased risk.
We also give explicit operator-block duals for the NH and Holevo bounds and identify the Holevo dual as a restriction of the NH dual.

The numerical applications examine finite-copy qubit models and full eight-parameter qutrit models.
We compare candidate measurements with finite Vidal measurements~\cite{VidalMixed,LPT}, and combine certified all-POVM risk lower bounds with certified upper bounds on the NH relaxation value.
Their strict separation proves that no locally unbiased measurement and estimator can attain the NH bound in the specified models.
For the two qutrit examples, the certified relative gaps exceed $14\%$ and $11\%$.
At level $m=1$, the certified lower bounds in these examples agree with the best numerical measurement costs to relative differences of approximately $4.5\times10^{-7}$ and $2.2\times10^{-6}$, indicating that a low level of the hierarchy can already yield numerically near-tight risk bounds.
The experiments use the fixed-coefficient hierarchy at level $m=1$; they do not numerically implement the joint-coefficient hierarchy in its full generality.
We retain this distinction, as well as the distinction between certified bounds and floating-point candidate risks, throughout the numerical analysis.

Section~\ref{sec:problem} specifies the estimation problem, its Fisher-information formulation, and the relation between measurement design and certification.
Section~\ref{sec:barc} describes the measurement search.
Sections~\ref{sec:setup}--\ref{sec:joint} develop the exact dual, the SOS hierarchy, and joint coefficient optimization, including reductions by sufficient subalgebras.
Section~\ref{sec:demo} presents the qubit and qutrit results, followed by the optimization benchmarks and timings, and Sec.~\ref{sec:conclusion} discusses their implications and scope.
Appendices~\mbox{\ref{app:nh}--\ref{app:finite-vidal}} provide the NH and Holevo dual formulas, the outcome bound for each algebra block, and the finite Vidal measurements used in the comparison.
Appendices~\ref{app:search-details} and \ref{app:hierarchy-proofs} contain the implementation details and the technical proofs for the SOS hierarchy.
\section{Optimization of locally unbiased estimation}\label{sec:problem}
\subsection{Quantum models, measurements, and estimators}\label{sec:model}
Let $\mathcal H$ be an $n$-dimensional complex Hilbert space. We denote by $\mathcal B(\mathcal H)$, $\mathcal{B}_h(\mathcal H)$, and $\mathcal S(\mathcal H)$ the space of linear operators, the real vector space of self-adjoint operators, and the set of density operators, respectively.
We write $\tr$ for the trace of an operator and $\Tr$ for the trace of a matrix on the parameter space.
The space of real symmetric $d\times d$ matrices is denoted by $\mathcal{B}_{hr}^d$.
Consider a local coordinate representation of a $d$-dimensional quantum statistical manifold on $\mathcal H$,
\[
 \{\rho_\theta\in\mathcal S(\mathcal H):\theta=(\theta^1,\ldots,\theta^d)\in\Theta\subset\R^d\}.
\]
We assume that $\Theta$ is an open coordinate domain and that $\theta\mapsto\rho_\theta$ is smooth.
Writing $\partial_j:=\partial/\partial\theta^j$, we impose throughout the paper the regularity assumption that
$\partial_1\rho_\theta,\ldots,\partial_d\rho_\theta$ are linearly independent over $\R$ at every $\theta\in\Theta$.
The estimation precision is evaluated at a fixed point $\theta_0\in\Theta$.

For an outcome number $s\in\mathbb N$, let $\Omega_s=\{1,\ldots,s\}$.
A measurement described by a positive operator-valued measure (POVM) is a family $M=(M_x)_{x\in\Omega_s}$ satisfying $M_x\ge0$ and $\sum_{x\in\Omega_s}M_x=I$. We denote the set of such measurements by $\mathcal M(\mathcal H,s)$.
The probability of outcome $x$ in state $\rho_\theta$ is $p_\theta(x)=\tr(\rho_\theta M_x)$.
An estimator consists of a measurement and a classical estimation map,
\[
 (M,\hat\theta),\qquad M\in\mathcal M(\mathcal H,s),\qquad
 \hat\theta:\Omega_s\longrightarrow\R^d.
\]
The expectation of its component $\hat\theta^i$ is
\[
 E_\theta[M,\hat\theta^i]=\sum_{x\in\Omega_s}\hat\theta^i(x)\tr(\rho_\theta M_x).
\]
The pair $(M,\hat\theta)$ is \emph{locally unbiased} at $\theta_0$ if
\begin{equation}\label{eq:lu}
 \begin{aligned}
 \sum_x\hat\theta^i(x)\tr(\rho_{\theta_0}M_x)&=\theta_0^i,\\
 \sum_x\hat\theta^i(x)\tr(\partial_j\rho_{\theta_0}M_x)&=\delta^i_j
 \end{aligned}
\end{equation}
for all $i,j=1,\ldots,d$.
Its covariance matrix, equivalently its mean-square-error matrix at the evaluation point, is
\begin{equation}\label{eq:variance}
 V_{\theta_0}[M,\hat\theta]^{ij}
 =\sum_x\bigl(\hat\theta^i(x)-\theta_0^i\bigr)
          \bigl(\hat\theta^j(x)-\theta_0^j\bigr)\tr(\rho_{\theta_0}M_x).
\end{equation}
For a positive definite weight $W\in\mathcal{B}_{hr}^d$, we define the optimal risk by
\begin{equation}\label{eq:risk}
 c_{\theta_0}(W)=\inf\left\{\Tr\bigl(WV_{\theta_0}[M,\hat\theta]\bigr):
 (M,\hat\theta)\text{ is locally unbiased at }\theta_0,\ s\in\mathbb N\right\}.
\end{equation}
The infimum is over all finite-outcome measurements and locally unbiased estimation maps.

\subsection{Local models and Fisher information}\label{sec:local-model}
Let $P=\operatorname{supp}\rho_{\theta_0}$ be the orthogonal projection onto the support $\ran\rho_{\theta_0}=(\ker\rho_{\theta_0})^\perp$, and let $Q=I-P$ be the projection onto the kernel. Then
\begin{equation}\label{eq:assumptions}
 \begin{gathered}
 \rho_{\theta_0}\ge0,\quad\tr\rho_{\theta_0}=1,\quad
 \partial_j\rho_{\theta_0}=(\partial_j\rho_{\theta_0})^{\dagger},\\
 \tr(\partial_j\rho_{\theta_0})=0,\qquad Q(\partial_j\rho_{\theta_0})Q=0.
 \end{gathered}
\end{equation}
Indeed, for $\ket v\in\ker\rho_{\theta_0}$,
the nonnegative function $\bra v\rho_\theta\ket v$ attains its minimum at $\theta_0$. Differentiating and applying the polarization identity gives the last condition.
Models with parameter-dependent support are allowed. One-sided boundary derivatives that violate this condition are excluded.

We write the classical Fisher information matrix as
\begin{equation}\label{eq:fisher}
 \begin{aligned}
 F_{\theta_0}[M]&=\sum_xg_{\theta_0}[M_x],\\
 g_{\theta_0}[X]_{ij}&=
 \frac{\tr(\partial_i\rho_{\theta_0}X)\tr(\partial_j\rho_{\theta_0}X)}
 {\tr(\rho_{\theta_0}X)}\qquad(X\ge0).
 \end{aligned}
\end{equation}
If the denominator vanishes, then $X=QXQ$, and Eq.~\eqref{eq:assumptions} implies that both derivative factors in the numerator vanish. We set $g_{\theta_0}[X]=0$ in this case.
Information is defined here through first derivatives at the evaluation point; no additional contribution from second-order changes in zero probabilities is included.
For example, outside our regularity assumptions, the model $\rho_\theta=\diag(1-\theta^2,\theta^2)$ has zero derivative at $\theta_0=0$, whereas the Fisher information of its eigenbasis measurement tends to $4$ as $\theta$ approaches zero. This limit differs from Eq.~\eqref{eq:fisher}.

The symmetric logarithmic derivative (SLD) operators $L_j=L_j^{\dagger}$ and the SLD Fisher information at the evaluation point are defined by
\begin{equation}\label{eq:sld-information}
 \partial_j\rho_{\theta_0}=\frac{\rho_{\theta_0}L_j+L_j\rho_{\theta_0}}2,\qquad
 (J_{\theta_0})_{ij}=\tr\bigl((\partial_i\rho_{\theta_0})L_j\bigr)
 =\operatorname{Re}\tr(\rho_{\theta_0}L_iL_j).
\end{equation}
The conditions in Eq.~\eqref{eq:assumptions} ensure that SLDs exist even for singular states.
Their only nonuniqueness is $L_j\mapsto L_j+K$, with $K=QKQ=K^{\dagger}$, which leaves $\partial_j\rho_{\theta_0}$ and $J_{\theta_0}$ unchanged.
Thus the SLDs and their metric are well defined on the quotient by operators supported on the kernel.
The SLD metric does not increase when the same quantum channel acts on the state and its tangent vectors.
In particular, the SLD Fisher information of the model obtained through the parameter-independent measurement channel
\[
 T^{(M)}(A)=\diag\bigl(\tr(AM_x)\bigr)_{x\in\Omega_s}
\]
is precisely the classical Fisher information $F_{\theta_0}[M]$ of $T^{(M)}(\rho_\theta)$. Monotonicity therefore gives
\begin{equation}\label{eq:sld-measurement}
 0\le F_{\theta_0}[M]\le J_{\theta_0}
\end{equation}
as in Ref.~\cite{YamagataMD}.
Although the monotone metrics in that reference are initially defined for positive definite states, the measurement inequality extends to the singular states considered here by using
$\rho_{\theta,\varepsilon}=(1-\varepsilon)\rho_\theta+\varepsilon I_n/n$ with $0<\varepsilon<1$.
Its derivatives are $(1-\varepsilon)\partial_j\rho_{\theta_0}$. The kernel-block condition in Eq.~\eqref{eq:assumptions} gives
$J_{\theta_0,\varepsilon}\to J_{\theta_0}$, while the vanishing derivatives of zero-probability outcomes give
$F_{\theta_0,\varepsilon}[M]\to F_{\theta_0}[M]$.
After removing zero effects $M_x=0$, we apply the measurement inequality to the positive definite model and take $\varepsilon\downarrow0$ to obtain Eq.~\eqref{eq:sld-measurement}.

Under the regularity assumption of Sec.~\ref{sec:model}, a measurement with positive definite Fisher information can be constructed by uniformly randomizing the projection-valued measures (PVMs) of orthonormalized SLDs.

\begin{proposition}[Randomization of SLD measurements]\label{prop:sld-randomization}
For the quantum statistical manifold considered here, there exists a finite-outcome POVM $M$ such that
\begin{equation}\label{eq:sld-randomization-bounds}
 \frac1dJ_{\theta_0}\le F_{\theta_0}[M]\le J_{\theta_0}.
\end{equation}
In particular, $F_{\theta_0}[M]>0$, and for every $W>0$,
$c_{\theta_0}(W)\le d\Tr(WJ_{\theta_0}^{-1})<\infty$.
\end{proposition}
\begin{proof}
The matrix $J:=J_{\theta_0}$ is the Gram matrix of the tangent vectors $\partial_1\rho_{\theta_0},\ldots,\partial_d\rho_{\theta_0}$ in the SLD metric, so $J>0$ by their real linear independence assumed in Sec.~\ref{sec:model}.
We therefore orthonormalize the SLDs as
\[
 \widetilde L_j=\sum_{i=1}^d(J^{-1/2})_{ji}L_i,\qquad
 \operatorname{Re}\tr(\rho_{\theta_0}\widetilde L_i\widetilde L_j)=\delta_{ij}.
\]
Let $\widetilde L_j=\sum_a\lambda_{ja}E_{ja}$ be the spectral decomposition,
and fix the PVM $E^{(j)}=(E_{ja})_a$ at the evaluation point.
Choosing each PVM with probability $1/d$ and recording both the choice $j$ and the outcome $a$ gives the finite-outcome POVM
\begin{equation}\label{eq:sld-randomization}
 M_{(j,a)}=\frac1dE_{ja},\qquad
 F_{\theta_0}[M]=\frac1d\sum_{j=1}^dF_{\theta_0}[E^{(j)}].
\end{equation}
Define the normalized Fisher matrices and tangent vectors by
\[
 \widehat F_j=J^{-1/2}F_{\theta_0}[E^{(j)}]J^{-1/2},\qquad
 \widehat D_j=\sum_i(J^{-1/2})_{ji}\partial_i\rho_{\theta_0}.
\]
Monotonicity~\eqref{eq:sld-measurement} gives $0\le\widehat F_j\le I_d$.
On the other hand, writing $p_{ja}=\tr(\rho_{\theta_0}E_{ja})$, we have
$\tr(\widehat D_jE_{ja})=\lambda_{ja}p_{ja}$, and therefore, including zero-probability outcomes,
\[
 (\widehat F_j)_{jj}=\sum_a p_{ja}\lambda_{ja}^2
 =\tr(\rho_{\theta_0}\widetilde L_j^2)=1.
\]
Let $e_j$ be the $j$th standard basis vector. Since the $j$th diagonal entry of $I_d-\widehat F_j\ge0$ vanishes,
$(I_d-\widehat F_j)e_j=0$, which implies $\widehat F_j\ge e_je_j^{\mathsf T}$.
Thus
\[
 J^{-1/2}F_{\theta_0}[M]J^{-1/2}
 =\frac1d\sum_j\widehat F_j\ge\frac1d I_d.
\]
The upper bound also follows from monotonicity~\eqref{eq:sld-measurement}.
Consequently, $F_{\theta_0}[M]^{-1}\le dJ^{-1}$. The risk bound follows by using the estimator attaining equality in Proposition~\ref{prop:cr} below.
\end{proof}

\subsection{Attaining the classical Cram\'er--Rao bound for a fixed measurement}
\begin{proposition}\label{prop:cr}
If $F_{\theta_0}[M]>0$, every estimator $(M,\hat\theta)$ that is locally unbiased at $\theta_0$ satisfies
\[
 V_{\theta_0}[M,\hat\theta]\ge F_{\theta_0}[M]^{-1}\ge J_{\theta_0}^{-1}.
\]
Equality in the first inequality can be attained by a suitable estimation map that is locally unbiased at $\theta_0$. Hence
\begin{equation}\label{eq:risk-fisher}
 c_{\theta_0}(W)=\inf_{s\in\mathbb N}\inf_{M\in\mathcal M(\mathcal H,s)}
 \Tr\bigl(WF_{\theta_0}[M]^{-1}\bigr).
\end{equation}
For singular Fisher matrices, the objective on the right is defined to be $+\infty$.
\end{proposition}
\begin{proof}
For $p_x=\tr(\rho_{\theta_0}M_x)>0$, define the score vector by
\[
 \ell_{\theta_0}(x)=\left(\frac{\tr(\partial_j\rho_{\theta_0}M_x)}{p_x}\right)_{j=1}^d,
\]
and set it to zero when $p_x=0$. Abbreviate $F=F_{\theta_0}[M]$ in this proof.
Expanding with the local unbiasedness conditions~\eqref{eq:lu} gives
\begin{equation}\label{eq:classical-square}
 \begin{aligned}
 V_{\theta_0}[M,\hat\theta]-F^{-1}
 &=\sum_{x:p_x>0}p_x\,r_xr_x^{\mathsf T}\ge0,\\
 r_x&=\hat\theta(x)-\theta_0-F^{-1}\ell_{\theta_0}(x).
 \end{aligned}
\end{equation}
The corresponding inequality also holds after taking the weighted trace.
The second inequality follows from SLD monotonicity~\eqref{eq:sld-measurement} and the order-reversing property of inversion.
Conversely, choose
\begin{equation}\label{eq:efficient-estimator}
 \hat\theta^i(x)=\theta_0^i+
 \sum_{j=1}^d\bigl(F_{\theta_0}[M]^{-1}\bigr)^{ij}
 \frac{\tr(\partial_j\rho_{\theta_0}M_x)}{\tr(\rho_{\theta_0}M_x)}
\end{equation}
for positive-probability outcomes, and set $\hat\theta(x)=\theta_0$ otherwise.
The expectation is $\theta_0$ because $\sum_x\tr(\partial_j\rho_{\theta_0}M_x)=0$. The definition of Fisher information gives the derivative constraints. All $r_x$ in Eq.~\eqref{eq:classical-square} vanish, yielding equality.
If $F$ is singular, a nonzero vector $c$ with $Fc=0$ is orthogonal to the derivative vector of every outcome, so the derivative constraints in Eq.~\eqref{eq:lu} cannot be satisfied.
\end{proof}
This attainability statement allows $\hat\theta$ to depend on the evaluation point $\theta_0$. It does not assert that one estimation map is efficient at every $\theta$.

\subsection{Rank-one measurements and a sufficient number of outcomes}\label{sec:rank-one-outcomes}
Let $M'=(M'_{x,a})$ be the POVM obtained by spectrally refining each effect $M_x$.
The original measurement satisfies $M_x=\sum_aM'_{x,a}$ and is obtained by the classical channel that discards $a$ from the refined outcome $(x,a)$.
Since the SLD metric on diagonal states equals the classical Fisher metric, monotonicity under this channel gives
$F_{\theta_0}[M]\le F_{\theta_0}[M']$.
Inversion reverses this order, so restricting to rank-one POVMs preserves the optimal value in Eq.~\eqref{eq:risk-fisher}.
Yamagata's outcome bound~\cite{Yamagata} states that, for any finite-outcome POVM $M$, there exists a POVM $M'$ with rank-one nonzero effects, at most
\begin{equation}\label{eq:outcomes}
 s_{\max}=(\dim\mathcal H)^2+\frac{d(d+1)}2-1
\end{equation}
outcomes, and $F_{\theta_0}[M']\ge F_{\theta_0}[M]$.
More generally, consider a real linear subspace $\mathcal A\subset\mathcal B(\mathcal H)$ containing $I$, with self-adjoint part $\mathcal A_h:=\mathcal A\cap\mathcal{B}_h(\mathcal H)$.
We call $\mathcal A$ a locally sufficient space at $\theta_0$ if there exists a positive real linear map $\Gamma:\mathcal{B}_h(\mathcal H)\to\mathcal A_h$ on the self-adjoint operator spaces relevant to measurements such that
\[
 \Gamma(I)=I,\qquad
 \tr(S\Gamma(H))=\tr(SH)
 \quad\bigl(S\in\{\rho_{\theta_0},\partial_1\rho_{\theta_0},\ldots,\partial_d\rho_{\theta_0}\},\ H\in\mathcal{B}_h(\mathcal H)\bigr).
\]
For measurements whose effects belong to $\mathcal A_h$, the term $(\dim\mathcal H)^2$ in the outcome bound can then be replaced by $\dim_{\R}\mathcal A_h$.
For general $\mathcal A$, the effects need not have rank one on $\mathcal H$; Eq.~\eqref{eq:outcomes} corresponds to $\mathcal A=\mathcal B(\mathcal H)$.
The construction for locally sufficient spaces that are algebras is described in Sec.~\ref{sec:barc-sufficiency}.

For the full operator space, Eq.~\eqref{eq:outcomes} ensures both rank-one effects and a bound on the number of outcomes, so restricting the search to $s\le s_{\max}$ preserves the infimum.
Zero effects may be added to equalize outcome numbers. Preserving the optimum with fewer outcomes requires a separate argument.
The infimum is attained for positive definite states, but a finite infimum may fail to be attained for singular states~\cite{Yamagata}.
We therefore formulate the problem in terms of its infimum and use a restriction to rank-one POVMs that preserves its value.

\subsection{Measurement design and independent certification}\label{sec:framework-overview}
Given the local data $\rho_{\theta_0},\partial_1\rho_{\theta_0},\ldots,\partial_d\rho_{\theta_0}$ and the weight $W$, the computation has two branches.
The nonconvex search in Sec.~\ref{sec:barc} optimizes a rank-one measurement and constructs its locally unbiased estimation map using Eq.~\eqref{eq:efficient-estimator}.
Independently, the dual formulation in Sec.~\ref{sec:setup} and the matrix-SOS hierarchy in Sec.~\ref{sec:hierarchy} produce a certified lower bound $\underline c_{\mathrm{cert}}$ valid for all measurements.
For any exactly feasible measurement $M$ with $F_{\theta_0}[M]>0$, these quantities satisfy
\begin{equation}\label{eq:search-certificate-overview}
 \underline c_{\mathrm{cert}}\le c_{\theta_0}(W)\le\Tr\bigl(WF_{\theta_0}[M]^{-1}\bigr).
\end{equation}
The separation between a lower certificate and an attainable risk therefore measures the remaining room for improvement.
Joint optimization of the support coefficients connects the NH bound to the optimal risk through a convergent hierarchy, as established in Sec.~\ref{sec:joint}.
The numerical experiments use fixed coefficients and distinguish the certified lower bound from the floating-point evaluation of the candidate risk; their certification conventions are specified in Sec.~\ref{sec:experiment-conventions}.
\section{Measurement search in Q-StARC}\label{sec:barc}
The measurement search in Q-StARC combines variable reduction by rank-one factorization, direct updates of complex Stiefel matrices, an initial Adam search, and refinement by the Riemannian conjugate gradient (RCG) method.
The computational savings arise from incorporating positivity and completeness into the measurement representation, so that objective evaluation and constraint handling reduce to matrix operations on smaller arrays.
Componentwise adaptation in Adam and a thin QR retraction reduce the cost per iteration, while RCG updates satisfying a descent condition further decrease the objective.
This section specifies the search algorithm and its computational cost. Section~\ref{sec:barc-sufficiency} describes an extension based on a sufficient algebra.

\subsection{Rank-one factorization and direct Stiefel representation}\label{sec:barc-representation}
For $s\ge n$ outcomes, we use the complex Stiefel manifold
\[
 \mathrm{St}_{\C}(s,n)=\{Z\in\C^{s\times n}:Z^{\dagger} Z=I_n\},
\]
the set of matrices with orthonormal columns. If $\ket{u_x}$ is the $x$th column of $Z^{\dagger}$, the effects $M_x=\ket{u_x}\bra{u_x}$ automatically satisfy completeness.
Conversely, every rank-one POVM has this representation. The search updates $Z$ directly.
Positivity is built into $M_x=\ket{u_x}\bra{u_x}$, and completeness is expressed by the single orthogonality condition $Z^{\dagger}Z=I_n$.
Thus the reduction to rank-one measurements in Sec.~\ref{sec:rank-one-outcomes} directly reduces the number of search variables.

For a fixed outcome number $s$, general Hermitian effects require $sn^2$ real variables and have $(s-1)n^2$ degrees of freedom in the interior after imposing completeness.
Measurement vectors require only $2sn$ real numbers, and the Stiefel manifold has real dimension $2sn-n^2$.
On the subset with no zero rows, the phase of each $\ket{u_x}$ leaves the measurement unchanged, so the corresponding POVMs have $2sn-n^2-s$ degrees of freedom after identifying phases.
For example, when $n=16,s=24$, general effects have $5888$ degrees of freedom, compared with $512$ for the Stiefel representation and $488$ after identifying phases.
We optimize on the Stiefel manifold while retaining this phase redundancy.
The difference in degrees of freedom between general effects and the Stiefel representation is $sn(n-2)$, giving a substantial reduction when $n>2$, as in multicopy systems.
A sufficient outcome number is chosen separately according to Sec.~\ref{sec:rank-one-outcomes}.

\subsection{Direct differentiation and feasibility via thin QR}\label{sec:barc-qr}
Let the current iterate be $Z\in\mathrm{St}_{\C}(s,n)$.
Regarding the ambient space $\C^{s\times n}$ as a real vector space, define the real Frobenius inner product by
\[
 \langle Y,T\rangle:=\operatorname{Re}\tr(Y^{\dagger}T)
\]
for $Y,T\in\C^{s\times n}$. The corresponding norm is denoted by $\|\cdot\|_{\mathrm F}$. The tangent space at $Z$ is
\[
 \mathrm T_Z\mathrm{St}_{\C}(s,n)
 :=\{T\in\C^{s\times n}:Z^{\dagger}T+T^{\dagger}Z=0\},
\]
and the restriction of this inner product to the tangent space defines the Riemannian metric.
For a square matrix $C\in\C^{n\times n}$, write its Hermitian part as
$\operatorname{sym}(C):=(C+C^{\dagger})/2$.
The orthogonal projection from the ambient space onto the tangent space is then
\[
 \Pi_Z:\C^{s\times n}\longrightarrow\mathrm T_Z\mathrm{St}_{\C}(s,n),\qquad
 \Pi_Z(Y):=Y-Z\operatorname{sym}(Z^{\dagger}Y)
 \quad(Y\in\C^{s\times n}).
\]
The gradient projection therefore requires only matrix products.

The Adam stage (Sec.~\ref{sec:barc-adam}) and the RCG stage (Sec.~\ref{sec:barc-rcg}) use the same thin QR retraction.
For a full-column-rank matrix $A\in\C^{s\times n}$, define $\operatorname{qf}(A):=Q$ by the unique factorization
\[
 \begin{gathered}
 A=QR,\qquad Q\in\C^{s\times n},\quad Q^{\dagger}Q=I_n,\qquad R\in\C^{n\times n},\\
 R_{ij}=0\quad(i>j),\qquad R_{ii}>0\quad(1\le i\le n).
 \end{gathered}
\]
For $T\in\mathrm T_Z\mathrm{St}_{\C}(s,n)$, set
\[
 \mathcal R_Z:\mathrm T_Z\mathrm{St}_{\C}(s,n)\longrightarrow\mathrm{St}_{\C}(s,n),\qquad
 \mathcal R_Z(T):=\operatorname{qf}(Z+T).
\]
This retraction preserves the first-order update direction while restoring completeness to numerical precision~\cite{AMS}.
The phase adjustment that defines these QR factors and verification of the retraction properties are given in Appendix~\ref{app:search-qr}.
The projection and thin QR each cost $O(sn^2)$, without requiring the linear systems of a general equality-constrained solver or tuning of penalty parameters.

Writing the measurement as $M(Z)=(\ket{u_x}\bra{u_x})_{x\in\Omega_s}$, we evaluate
\[
 p_x(Z)=\bra{u_x}\rho_{\theta_0}\ket{u_x},\qquad
 \dot p_x(Z)=\bigl(\bra{u_x}\partial_j\rho_{\theta_0}\ket{u_x}\bigr)_{j=1}^d
\]
directly. Using the same quadratic forms on the ambient space $\C^{s\times n}$, define
\[
 F(Z):=\sum_{x\in\Omega_s}\frac{\dot p_x(Z)\dot p_x(Z)^{\mathsf T}}{p_x(Z)},\qquad
 f_W(Z):=\Tr\bigl(WF(Z)^{-1}\bigr).
\]
The differentiation formulas below apply on the open region where all $p_x(Z)>0$ and $F(Z)>0$.
On the Stiefel manifold, $F(Z)=F_{\theta_0}[M(Z)]$, so $f_W$ is the measurement optimization objective.
Abbreviating $p_x=p_x(Z)$, $\dot p_x=\dot p_x(Z)$, and $F=F(Z)$, we obtain
\begin{equation}\label{eq:gradient}
 \begin{aligned}
 \mathrm df_W&=-\Tr(H_F\,\mathrm dF),\qquad H_F=F^{-1}WF^{-1},\\
 \mathrm dF&=\sum_x\left\{\frac{(\mathrm d\dot p_x)\dot p_x^{\mathsf T}
 +\dot p_x(\mathrm d\dot p_x)^{\mathsf T}}{p_x}
 -\frac{\dot p_x\dot p_x^{\mathsf T}}{p_x^2}\,\mathrm dp_x\right\}.
 \end{aligned}
\end{equation}
The Euclidean gradient $\nabla f_W(Z)\in\C^{s\times n}$ of this ambient extension is defined by
$\mathrm df_W(Z)[Y]=\langle\nabla f_W(Z),Y\rangle$ for $Y\in\C^{s\times n}$.
The search uses the Riemannian gradient
\[
 g(Z):=\operatorname{grad}f_W(Z)
 =\Pi_Z(\nabla f_W(Z))\in\mathrm T_Z\mathrm{St}_{\C}(s,n).
\]
Following Yamagata's master's thesis, Kimizu \emph{et al.}~\cite{KTF} parameterized isometries representing real rank-one POVMs by products of Givens rotations, termed two-level orthogonal matrices in that work, and optimized them using basin hopping and Powell's method.
A Givens rotation $G_{ab}(\varphi)$ acts on two selected real coordinates $a,b$ as
\[
 \begin{pmatrix}\cos\varphi&-\sin\varphi\\ \sin\varphi&\cos\varphi\end{pmatrix}
 \qquad(\varphi\in\R)
\]
and fixes all remaining coordinates. A product representation uses the rotation angles as search variables.
Both approaches restrict to rank-one measurements; Q-StARC combines direct updates of the Stiefel matrix $Z$ with Adam and a QR retraction.
Direct differentiation avoids constructing $Z$ from rotation angles and propagating derivatives through that construction.
Automatic differentiation is applied to $f_W(Z)$ at the current iterate, without backpropagating through the QR update or the full optimization history.

\subsection{First stage: projected ambient Adam}\label{sec:barc-adam}
We apply the componentwise adaptation based on Adam's first and second moments~\cite{Adam} to the real and imaginary parts of the measurement matrix.
Define the real-coordinate map $\mathcal P:\C^{s\times n}\to\R^{2\times s\times n}$ by
$\mathcal P(Y)=(\operatorname{Re}Y,\operatorname{Im}Y)$.
Its inverse is $\mathcal P^{-1}(A,B)=A+iB$ for $A,B\in\R^{s\times n}$.
Starting from $Z_1\in\mathrm{St}_{\C}(s,n)$, initialize the real first- and second-moment arrays $m_t,v_t\in\R^{2\times s\times n}$ with $m_0=v_0=0$.
At iteration $t=1,2,\ldots$, set $g_t=g(Z_t)$ and $a_t=\mathcal P(g_t)$, and update
\begin{align}
 m_t&=\beta_1m_{t-1}+(1-\beta_1)a_t,&
 v_t&=\beta_2v_{t-1}+(1-\beta_2)a_t^{\odot2},\nonumber\\
 \widehat m_t&=m_t/(1-\beta_1^t),&
 \widehat v_t&=v_t/(1-\beta_2^t),\nonumber\\
 T_t&=-\eta\Pi_{Z_t}\mathcal P^{-1}
       \left(\widehat m_t\oslash(\sqrt{\widehat v_t}+\epsilon)\right),&
 Z_{t+1}&=\mathcal R_{Z_t}(T_t).
\end{align}
The superscript $\odot2$ denotes the componentwise square, $\oslash$ denotes componentwise division, and the square root is taken componentwise; $\epsilon$ is added to every component.
The parameters $0\le\beta_1,\beta_2<1$ are the moment decay rates, $\eta>0$ is the learning rate, and $\epsilon>0$ prevents division by zero.
Their default values and the componentwise conventions are specified in Appendix~\ref{app:search-adam}.
A second projection is required because componentwise rescaling can move the update direction outside the tangent space.
The Adam stage uses a fixed learning rate and requires no additional objective evaluations for line-search trials.
Updating the componentwise history takes $O(sn)$ time and memory and adjusts the update scales without storing a large metric matrix or an inverse-Hessian approximation.
The iterate after the prescribed number of updates initializes RCG.

Because $m_t,v_t$ are not transported between tangent spaces, this method is \emph{projected ambient Adam}.
Its componentwise updates depend on ambient coordinates and therefore differ from a coordinate-invariant geometric Riemannian Adam method.

\subsection{Second stage: RCG with PR+ and Armijo line search}\label{sec:barc-rcg}
Relabel the final Adam iterate as $Z_0$ and index RCG iterations by $t=0,1,\ldots$.
Let $g_t=g(Z_t)$ and $\Delta_t\in\mathrm T_{Z_t}\mathrm{St}_{\C}(s,n)$ denote the gradient and search direction, initialized by $\Delta_0=-g_0$.
After an accepted step $\alpha_t>0$ gives $Z_{t+1}=\mathcal R_{Z_t}(\alpha_t\Delta_t)$, compute $g_{t+1}=g(Z_{t+1})$.
Define the projection transport between consecutive tangent spaces by
\[
 \mathcal T_t:\mathrm T_{Z_t}\mathrm{St}_{\C}(s,n)\longrightarrow
 \mathrm T_{Z_{t+1}}\mathrm{St}_{\C}(s,n),\qquad
 \mathcal T_t(Y):=\Pi_{Z_{t+1}}(Y).
\]
The next search direction is
\[
 \Delta_{t+1}=-g_{t+1}+\beta_t^{\mathrm{PR+}}\mathcal T_t\Delta_t,
\]
where $\beta_t^{\mathrm{PR+}}\ge0$ is the PR+ coefficient defined in Eq.~\eqref{eq:rcg}.
We select the accepted step $\alpha_t$ by Armijo backtracking and reset the direction to steepest descent whenever it fails the descent condition.
The precise PR+ coefficient, line search, and stopping rules are given in Appendix~\ref{app:search-rcg}.
Only the objective is evaluated at trial points; a new gradient is computed at the accepted point, avoiding gradient evaluations for rejected step sizes.

\subsection{Cost of objective evaluation and iteration}\label{sec:barc-cost}
Objective evaluation uses batched contractions of the quadratic forms in $\ket{u_x}$, without explicitly constructing the $s$ effect matrices of size $n\times n$.
The memory required for the measurement representation is therefore reduced from $O(sn^2)$ to $O(sn)$.
The Fisher matrix is indexed by the $d$ estimated parameters, independently of the Hilbert-space dimension $n$, so objective evaluation requires a $d\times d$ linear system.
For example, estimating the three components of a multicopy qubit model has $d=3$, regardless of the number of copies.
After reduction to unit weight, objective evaluation checks positive definiteness by Cholesky factorization and computes $\Tr F^{-1}$ by solving linear systems with coefficient matrix $F$.

Table~\ref{tab:barc-cost} lists the main operation counts for dense model matrices.

\begin{table}[t]
\centering
\caption{Operation counts for the Q-StARC measurement search. Projection and QR costs are per application.}\label{tab:barc-cost}
\begin{tabular}{@{}ll@{}}
\toprule
Operation & Cost\\
\midrule
Outcome probabilities and $d$ derivatives & $O((d+1)sn^2)$\\
Fisher matrix assembly & $O(sd^2)$\\
Fisher linear solve & $O(d^3)$\\
Tangent projection or thin QR & $O(sn^2)$ each\\
Adam moment update & $O(sn)$\\
\bottomrule
\end{tabular}
\end{table}

An Adam iteration requires gradient evaluation, two projections, and a QR factorization. RCG additionally requires objective evaluations and QR factorizations according to the number of line-search trials.
In addition to the measurement representation, storage is required for the model, of size $O((d+1)n^2)$, and for intermediate automatic-differentiation arrays.

The implementation uses JAX automatic differentiation and just-in-time (JIT) compilation, with the Adam iterations, RCG iterations, and Armijo search executed in compiled loops.
Executing multiple iterations together reduces per-iteration Python calls and synchronization. Compiled computations are reused when compilation conditions, including array shapes, data types, and static arguments, are unchanged.
Rank-one factorization reduces measurement storage, direct Stiefel updates avoid constructing matrices from angular coordinates, and fixed-learning-rate Adam with thin QR reduces the cost per iteration.
In addition to these structural reductions, the overall performance is assessed by the comparison experiments in Sec.~\ref{sec:barc-evidence}.

\subsection{Blockwise optimization with a sufficient algebra}\label{sec:barc-sufficiency}
A sufficient algebra for local estimation allows further reduction of the search space while preserving the optimal risk~\cite{Yamagata,YamagataSuff}.
We formulate blockwise optimization as a theoretical extension of the measurement search. This extension is not implemented in the measurement-search component used in this paper and is not used in the numerical experiments.
Here we abbreviate $\rho:=\rho_{\theta_0}$ and $D_j:=\partial_j\rho_{\theta_0}$.
For a unital subalgebra $\mathcal A\subset\mathcal B(\mathcal H)$, let $\mathcal A_h:=\mathcal A\cap\mathcal{B}_h(\mathcal H)$ be its Hermitian part.
Sufficiency at $\theta_0$ means that there exists a positive unital real linear map
$\Gamma:\mathcal{B}_h(\mathcal H)\to\mathcal A_h$ satisfying
\begin{equation}\label{eq:barc-local-sufficiency}
 \Gamma(I)=I,\qquad
 \tr\bigl(S\Gamma(H)\bigr)=\tr(SH)
 \quad\left(S\in\{\rho,D_1,\ldots,D_d\},\ H=H^{\dagger}\right).
\end{equation}
Replacing $M_x$ by $\Gamma(M_x)$ therefore preserves the outcome probabilities, their first derivatives, and the Fisher information.

For simplicity, we consider a decomposition into complex matrix blocks in a fixed basis,
\begin{equation}\label{eq:barc-sufficient-blocks}
 \mathcal H=\bigoplus_{b=1}^{q}\left(\C^{n_b}\otimes\C^{m_b}\right),\qquad
 \mathcal A=\bigoplus_{b=1}^{q}\left(\mathbb M_{n_b}(\C)\otimes I_{m_b}\right),\qquad
 n=\sum_{b=1}^{q}n_bm_b.
\end{equation}
Here $q$ is the number of blocks, $m_b$ is the multiplicity, and $\mathbb M_{n_b}(\C)$ is the algebra of $n_b\times n_b$ complex matrices.
Define the embedding of block $b$ by $\iota_b:\C^{n_b}\otimes\C^{m_b}\to\mathcal H$, with $\iota_b(v)=(0,\ldots,0,v,0,\ldots,0)$, and form the partial traces
\[
 \rho_b:=\tr_{\C^{m_b}}(\iota_b^{\dagger}\rho\iota_b),\qquad
 D_{j,b}:=\tr_{\C^{m_b}}(\iota_b^{\dagger}D_j\iota_b)
 \quad(j=1,\ldots,d).
\]
They satisfy $\sum_b\tr\rho_b=1$. The matrices $\rho_b$ are left unnormalized so that the block probabilities and their derivatives are retained.

Following the notation of Ref.~\cite{Yamagata}, define
\[
 \begin{gathered}
 \mathcal A_+:=\{E\in\mathcal A_h:E\ge0\},\qquad
 \mathcal A_1:=\{E\in\mathcal A_+:\tr E\le1\},\\
 \mathcal A_e:=\{E\in\mathcal A_+:\exists a>0,\ aE\in\operatorname{ext}\mathcal A_1\}.
 \end{gathered}
\]
Here $\operatorname{ext}\mathcal A_1$ is the set of extreme points of $\mathcal A_1$.
In this block decomposition, the nonzero elements of $\mathcal A_e$ are positive multiples of minimal projections of $\mathcal A$ supported in a single block, and lie on extreme rays of the positive cone $\mathcal A_+$.
Refining a measurement in the algebra by its blocks and their spectral components cannot decrease Fisher information, by SLD monotonicity under classical coarse graining, as in Sec.~\ref{sec:rank-one-outcomes}.
Consequently, restricting effects to $\mathcal A_e$ preserves the infimum of the risk.
When the block index $b$ is retained as part of the outcome, these effects admit rank-one factors on each reduced block.
For outcome numbers $s_b\ge n_b$, write
\begin{equation}\label{eq:barc-product-stiefel}
 \begin{gathered}
 \boldsymbol Z=(Z_1,\ldots,Z_q)\in
 \prod_{b=1}^{q}\mathrm{St}_{\C}(s_b,n_b),\qquad
 \ket{u_{b,x}}:=Z_b^{\dagger}\ket{e_x}\quad(1\le x\le s_b),\\
 M_{b,x}:=\iota_b\bigl(\ket{u_{b,x}}\bra{u_{b,x}}\otimes I_{m_b}\bigr)\iota_b^{\dagger},\qquad
 \sum_{b=1}^{q}\sum_{x=1}^{s_b}M_{b,x}=I.
 \end{gathered}
\end{equation}
Here $\ket{e_x}$ is the standard basis of $\C^{s_b}$.
Each $M_{b,x}$ belongs to $\mathcal A_e$. For a nonzero effect, the reduced effect on $\C^{n_b}$ has rank one, whereas its rank on the original space $\mathcal H$ is $m_b$.
Applying Yamagata's outcome-elimination argument~\cite{Yamagata} to each block gives
\[
 n_b\le s_b\le s_b^{\max}:=n_b^2+\frac{d(d+1)}{2}-1
\]
without decreasing its Fisher information; see Theorem~\ref{thm:block-support} in Appendix~\ref{app:block-support}.
Measurements with fewer outcomes can be extended to $s_b=s_b^{\max}$ by splitting nonzero effects into positive multiples, preserving Fisher information.
Thus optimization on $\prod_b\mathrm{St}_{\C}(s_b^{\max},n_b)$ with the row count of each block fixed at its bound preserves the optimal-risk infimum, and no search over outcome allocations is required.

For each outcome, define its probability and derivative vector by
\[
 p_{b,x}=\bra{u_{b,x}}\rho_b\ket{u_{b,x}},\qquad
 \dot p_{b,x}=\bigl(\bra{u_{b,x}}D_{j,b}\ket{u_{b,x}}\bigr)_{j=1}^{d}.
\]
The objective to minimize is
\begin{equation}\label{eq:barc-block-fisher}
 F_b(Z_b):=\sum_{x:\,p_{b,x}>0}\frac{\dot p_{b,x}\dot p_{b,x}^{\mathsf T}}{p_{b,x}},\qquad
 F(\boldsymbol Z)=\sum_{b=1}^{q}F_b(Z_b),\qquad
 f_W(\boldsymbol Z)=\Tr\bigl(WF(\boldsymbol Z)^{-1}\bigr).
\end{equation}
The inverse is taken after summing all blocks, with $F(\boldsymbol Z)>0$. Gradients for all blocks are derived from this common objective and optimized jointly.
The product manifold is equipped with the sum of the block Frobenius metrics.
For $\boldsymbol Y=(Y_b)_b$, with $Y_b\in\C^{s_b\times n_b}$, and
$\boldsymbol T=(T_b)_b$, with $T_b\in\mathrm T_{Z_b}\mathrm{St}_{\C}(s_b,n_b)$, the projection and retraction are
\[
 \Pi_{\boldsymbol Z}(\boldsymbol Y)=\bigl(\Pi_{Z_b}(Y_b)\bigr)_{b=1}^{q},\qquad
 \mathcal R_{\boldsymbol Z}(\boldsymbol T)=\bigl(\operatorname{qf}(Z_b+T_b)\bigr)_{b=1}^{q}.
\]
Thus the Adam and RCG schemes above can be applied with blockwise projections and QR factorizations.

After preparing the reduced data, let $s:=\sum_b s_b$.
Storing the same POVM as $s$ dense $n\times n$ effect matrices would require $O(sn^2)$ memory and $O((d+1)sn^2)$ operations for probability and derivative traces.
The reduced block-factor representation instead requires $O(\sum_b s_bn_b)$ memory and $O((d+1)\sum_b s_bn_b^2)$ operations, respectively.
At each iteration, the projections and thin QR factorizations each cost $O(\sum_b s_bn_b^2)$, while model storage is $O((d+1)\sum_b n_b^2)$.
The real dimension of the search manifold is $\sum_b(2s_bn_b-n_b^2)$.
Fisher assembly and the linear solve retain their costs $O(sd^2)$ and $O(d^3)$, respectively, but matrix operations and memory can be substantially reduced when there are many small blocks or large multiplicities $m_b$.
These gains are based on operation and storage counts; no empirical speedup is evaluated for this extension.

\subsection{Outputs and numerical conventions}\label{sec:barc-weighted-implementation}
The search returns the measurement $Z$, Fisher matrix, risk, gradient norm, completeness residual, termination reason, and computation time for each stage.
With multiple initializations, the runs are independent and their admissible candidates are compared by risk.
Timing conventions, including treatment of the initial JIT compilation, are given in Sec.~\ref{sec:demo}. Optimality is assessed by comparison with the SOS lower bounds developed below.

The search routine uses the unit-weight objective $W=I_d$.
Here we abbreviate $D_j:=\partial_j\rho_{\theta_0}$ and $F:=F_{\theta_0}[M]$.
For general $W>0$, the local coordinate transformation $\theta-\theta_0=W^{-1/2}\eta$, with $\eta\in\R^d$, gives
\[
 \widetilde D_i:=\sum_j(W^{-1/2})_{ji}D_j,\qquad
 \widetilde F=W^{-1/2}FW^{-1/2},\qquad
 \Tr\widetilde F^{-1}=\Tr(WF^{-1}),
\]
reducing the problem to unit weight.
Sections~\ref{sec:experiment3} and \ref{sec:experiment4} handle general weights by passing these transformed derivatives to the search routine.
Iteration limits and learning rates for individual experiments are listed in Sec.~\ref{sec:experiment-conventions}.
Numerically, a nonzero effect with $p_x\le0$ produces a not-a-number (NaN) objective; only zero rows are assigned zero contribution.
The measurement search therefore operates where all nonzero effects have positive probabilities, whereas the theory in the following sections applies to arbitrary ranks.
Small positive probabilities are not clipped, and no ridge term is added to the Fisher matrix, such as $F\mapsto F+\lambda I_d$ with $\lambda>0$.
\section{Upper bounds from candidate measurements and lower bounds from the Fisher support function}\label{sec:setup}
We continue to fix the point $\theta_0$. To simplify the dual formulas, write
\[
 \rho:=\rho_{\theta_0},\qquad D_j:=\partial_j\rho_{\theta_0}
\]
We retain the notation $F_{\theta_0}[M]$ for the Fisher information matrix,
abbreviating it to $F$ in identities involving a single matrix.
The auxiliary vector in the SOS formulation is denoted by $\xi\in\R^d$, to distinguish it from the number of outcomes $s$.
\subsection{Completing the square for a general weight}
Let $\widehat M$ be a feasible measurement produced by the measurement search in Q-StARC, with $\widehat F=F_{\theta_0}[\widehat M]>0$. Its risk $c_{\mathrm{meas}}=\Tr(W\widehat F^{-1})$ is an upper bound on $c_{\theta_0}(W)$.
A lower bound requires a separate argument. For $F>0$ and any real matrix $B\in\R^{d\times d}$, completing the square gives
\begin{equation}\label{eq:square}
 \Tr(WF^{-1})-2\Tr(BW^{1/2})+\Tr(BFB^{\mathsf T})
 =\|(B-W^{1/2}F^{-1})F^{1/2}\|_{\mathrm F}^2\ge0
\end{equation}
Defining the support function $h$ for $H\in\mathcal{B}_{hr}^d$, we obtain
\begin{equation}\label{eq:fixed-bound}
 h(H)=\sup_M\Tr(HF_{\theta_0}[M]),\quad h_B=h(B^{\mathsf T}B),\qquad
 h_B\le\overline h\ \Longrightarrow\ 2\Tr(BW^{1/2})-\overline h\le c_{\theta_0}(W).
\end{equation}
Here $h$ is the support function of the Fisher information set at $\theta_0$, with the supremum taken over all $s\in\mathbb N$ and $M\in\mathcal M(\mathcal H,s)$.
The coefficient that makes Eq.~\eqref{eq:square} an equality at the candidate is
\begin{equation}\label{eq:Bchoice}
 B=W^{1/2}\widehat F^{-1},\qquad B^{\mathsf T}B=\widehat F^{-1}W\widehat F^{-1}.
\end{equation}
In general, $B$ is not symmetric. We first keep this $B$ fixed.
Since $h_{tB}=t^2h_B$, if $\overline h>0$, an additional lower bound is available without solving another SDP:
\begin{equation}\label{eq:scaled-bound}
 \frac{\{\Tr(BW^{1/2})\}^2}{\overline h}\le c_{\theta_0}(W)
\end{equation}
This is the maximum of $2t\Tr(BW^{1/2})-t^2\overline h$ over $t\in\R$.

\subsection{A concave value function and supporting hyperplanes}\label{sec:concave-value}
To identify the meaning of the dual variable, we replace the right-hand side $I$ of the POVM completeness condition by an arbitrary positive semidefinite operator $S$.
Using $g_{\theta_0}[E]$ from Eq.~\eqref{eq:fisher}, define
\[
 \gamma_B(E):=\Tr\bigl(Bg_{\theta_0}[E]B^{\mathsf T}\bigr)
\]
for any positive semidefinite operator $E$. The convention $g_{\theta_0}[E]=0$ at zero probability also gives $\gamma_B(E)=0$ in that case.
The function $\gamma_B$ is nonnegative and positively homogeneous, and every POVM $M$ satisfies
$\sum_x\gamma_B(M_x)=\Tr(BF_{\theta_0}[M]B^{\mathsf T})$.
We therefore define
\[
 v(S):=\sup_{\substack{r\in\mathbb N,\ E_1,\ldots,E_r\ge0\\
                       \sum_{\alpha=1}^rE_\alpha=S}}
             \sum_{\alpha=1}^r\gamma_B(E_\alpha)
 \qquad(S\ge0)
\]
The supremum is over all finite decompositions, with no prescribed number of terms, and $v(I)=h_B$.

Monotonicity of the SLD metric, Eq.~\eqref{eq:sld-measurement}, implies
$h_B\le\Tr(BJ_{\theta_0}B^{\mathsf T})<\infty$.
For $S\ne0$, set $c:=\|S\|_{\mathrm{op}}>0$, where $\|\cdot\|_{\mathrm{op}}$ is the operator norm. For any decomposition $\sum_\alpha E_\alpha=S$,
adjoining $I-S/c$ to $\{E_\alpha/c\}_\alpha$ gives a POVM.
Positive homogeneity and nonnegativity of $\gamma_B$ yield $\sum_\alpha\gamma_B(E_\alpha)\le c h_B$, and hence
\[
 0\le v(S)\le\|S\|_{\mathrm{op}}h_B<\infty
\]
Since $v(0)=0$, the function $v$ is finite on the entire positive semidefinite cone.
Scaling decompositions and combining two decompositions that approximate their respective suprema to arbitrary accuracy give
\[
 v(tS)=t\,v(S)\quad(t\ge0),\qquad
 v(S_1+S_2)\ge v(S_1)+v(S_2)\quad(S_1,S_2\ge0)
\]
Thus $v$ is positively homogeneous and superadditive. In particular, it is concave:
\[
 v\bigl(tS_1+(1-t)S_2\bigr)
 \ge t\,v(S_1)+(1-t)v(S_2)\qquad(0\le t\le1)
\]

In general, if $f:\mathcal D\to\R$ is a finite concave function on a convex set $\mathcal D\subset\R^k$ and $x\in\operatorname{int}\mathcal D$, the existence of an upper supporting hyperplane gives
\[
 f(x)=\min_{u\in\R^k,\ b\in\R}
 \{u^{\mathsf T}x+b:u^{\mathsf T}y+b\ge f(y)\quad\forall y\in\mathcal D\}
\]
This representation selects an affine majorant of $f$ that agrees with $f$ at $x$; differentiability is not required.
Apply this fact to the concave function $v$ on the real vector space of Hermitian matrices, at the interior point $I$ of its domain.
Representing the upper supporting hyperplane by the trace inner product, there exists $X=X^{\dagger}$ such that
\[
 v(S)\le v(I)+\tr[X(S-I)]\qquad(S\ge0)
\]
This inequality also holds for boundary points $S$. Indeed, $(1-t)I+tS>0$ for $0<t<1$, so concavity and the supporting inequality at interior points imply
\[
 (1-t)v(I)+t\,v(S)
 \le v\bigl((1-t)I+tS\bigr)
 \le v(I)+t\,\tr[X(S-I)]
\]
The same inequality therefore follows without assuming continuity at the boundary.
Substituting $S=tI$ for both $0<t<1$ and $t>1$, positive homogeneity gives $\tr X=v(I)$.
The constant term of the supporting affine function consequently vanishes, and
\begin{equation}\label{eq:concave-support-dual}
 h_B=v(I)=\min_{X=X^{\dagger}}
 \{\tr X:\tr(XS)-v(S)\ge0\quad\forall S\ge0\}.
\end{equation}
Every feasible $X$ defines a linear majorant $S\mapsto\tr(XS)$ of $v$. An optimal $X$ is the slope of an upper supporting hyperplane at $I$, namely a \emph{supergradient} of $v$.
If $v$ is differentiable at $I$, this is its gradient $\nabla v(I)$ with respect to the trace inner product.
The supporting hyperplane ensures attainment of the minimum; attainment of the supremum over measurement decompositions defining $v$ is not needed.

This minimization representation can also be interpreted through Legendre--Fenchel duality for concave functions.
For $X=X^{\dagger}$, define the concave conjugate by
\[
 v_*(X):=\inf_{S\ge0}\{\tr(XS)-v(S)\}
\]
Positive homogeneity and $v(0)=0$ imply that $v_*(X)=0$ when $X$ satisfies the linear-majorant condition above, and $v_*(X)=-\infty$ otherwise.
Thus $v(I)=\min_{X=X^{\dagger}}\{\tr X-v_*(X)\}$ at the interior point $I$. The supporting-hyperplane argument directly establishes equality and attainment in this dual representation.

\subsection{Exact duality through a quadratic matrix polynomial}\label{sec:exact-dual}
The linear-majorant condition in the preceding subsection is equivalent to
\[
 \tr(XS)\ge v(S)\quad(\forall S\ge0)
 \quad\Longleftrightarrow\quad
 \tr(XE)\ge\gamma_B(E)\quad(\forall E\ge0)
\]
The forward implication follows from $\gamma_B(E)\le v(E)$, using the one-term decomposition. For the reverse implication, sum the inequalities over any finite decomposition and take the supremum.
To rewrite this constraint, introduce the Hermitian matrices $A_i:=\sum_{j=1}^dB_{ij}D_j$ ($i=1,\ldots,d$) and the notation
\[
 p[E]:=\tr(\rho E),\qquad
 a(E):=\bigl(\tr(A_iE)\bigr)_{i=1}^d\in\R^d
\]
If $p[E]>0$, Eq.~\eqref{eq:fisher} gives $\gamma_B(E)=\|a(E)\|^2/p[E]$. Completing the square yields
\[
 \gamma_B(E)=\sup_{\xi\in\R^d}
 \{2\xi^{\mathsf T}a(E)-p[E]\|\xi\|^2\}
\]
The representation also holds when $p[E]=0$, because then $a(E)=0$.
The linear-majorant condition is therefore equivalent to
\[
 \tr\!\left[\left(X+\|\xi\|^2\rho-2\sum_i\xi_iA_i\right)E\right]\ge0
\]
for every $E\ge0$ and $\xi\in\R^d$. By self-duality of the positive semidefinite cone, this is precisely the condition that the quadratic matrix polynomial
\begin{equation}\label{eq:Q}
 A(\xi):=\sum_i\xi_iA_i,\qquad
 Q_{B,X}(\xi):=X+\|\xi\|^2\rho-2A(\xi)
\end{equation}
be positive semidefinite for all $\xi\in\R^d$. We also write $Q_X$ when $B$ is fixed.
Thus $Q_{B,X}$ expresses the linear-majorant condition for the concave function $v$ as a matrix positivity condition.

The following theorem gives a fixed-coefficient Fisher-support formulation of the quantum-estimation duality established by Hayashi~\cite{Hayashi} and reformulated by Hayashi and Ouyang~\cite{HO}; the risk dual in those works also optimizes over the coefficients.
Here we give a direct alternative proof using the concave function $v(S)$ on the real $n^2$-dimensional space $\mathcal B_h(\mathcal H)$, whose supporting hyperplane at $I$ yields dual equality and attainment of the minimum.

\begin{theorem}[Exact duality for fixed coefficients]\label{thm:exact-dual}
Under assumptions~\eqref{eq:assumptions},
\[
 h_B=\min_{X=X^{\dagger}}\{\tr X:Q_{B,X}(\xi)\ge0\quad\forall \xi\in\R^d\}.
\]
The minimum on the right is attained, and $h_B\le\tr X$ for every feasible $X$.
Existence of an optimal measurement on the left is not required.
\end{theorem}
\begin{proof}
The result follows from the minimization representation of $v(I)$ in Sec.~\ref{sec:concave-value} and the equivalence of the constraints established above.
Existence of a minimizer follows from the existence of an upper supporting hyperplane at $I$.
\end{proof}

The upper bound from monotonicity of the SLD metric also has an explicit SOS certificate at the initial level; matrix SOS is defined in Sec.~\ref{sec:sos-definition}.
The SLDs $L_j$ can be chosen as Hermitian solutions of $D_j=(\rho L_j+L_j\rho)/2$.
In an eigenbasis of $\rho$ with eigenvalues $\lambda_a$, choose the matrix entry to be $2(D_j)_{ab}/(\lambda_a+\lambda_b)$ when $\lambda_a+\lambda_b>0$, and zero otherwise.
For $L_i^B=\sum_jB_{ij}L_j$ and $K_B=\sum_iL_i^B\rho L_i^B$, we have
\begin{equation}\label{eq:sld-sos}
 Q_{B,K_B}(\xi)=\sum_i(\xi_iI-L_i^B)\rho(\xi_iI-L_i^B)
\end{equation}
and $h_B\le\tr K_B=\Tr(BJ_{\theta_0}B^{\mathsf T})<\infty$.
The next section constructs a hierarchy of finite SDPs for the infinitely many positivity constraints.
\section{A denominator-based matrix-SOS hierarchy and convergence for arbitrary state rank}\label{sec:hierarchy}
\subsection{SOS representations and a hierarchy of finite SDPs}\label{sec:sos-definition}
An SOS representation certifies polynomial nonnegativity through a sum-of-squares identity.
For real variables $\xi=(\xi_1,\ldots,\xi_d)$, let $\R[\xi]$ and $\C[\xi]$ denote the polynomial rings with real and complex coefficients, respectively.
A real polynomial $p\in\R[\xi]$ is a \emph{scalar SOS} if it admits the identity
\[
 p(\xi)=\sum_{\nu=1}^N h_\nu(\xi)^2
 \qquad(N\in\mathbb N)
\]
for finitely many real polynomials $h_\nu\in\R[\xi]$. Such a representation implies $p(\xi)\ge0$ for every $\xi\in\R^d$.

We use the complex Hermitian matrix version of this notion.
An $n\times n$ Hermitian matrix polynomial $H(\xi)=H(\xi)^{\dagger}$ with entries in $\C[\xi]$ is a \emph{matrix SOS} if there are finitely many polynomial vectors $\ket{f_\nu}\in\C[\xi]^n$ such that
\[
 H(\xi)=\sum_{\nu=1}^N\ket{f_\nu(\xi)}\bra{f_\nu(\xi)}
 \qquad(\forall\xi\in\R^d)
\]
The variables $\xi$ are real, whereas the coefficients may be complex.
This is the usual definition for real symmetric matrices~\cite{DHL}, extended with this convention.
For every $\ket u\in\C^n$,
\[
 \bra uH(\xi)\ket u
 =\sum_{\nu=1}^N|\langle f_\nu(\xi)\mid u\rangle|^2\ge0
\]
so a matrix-SOS representation certifies positive semidefiniteness at every point.
The converse fails in general: a matrix polynomial that is positive semidefinite everywhere need not be a matrix SOS. A scalar counterexample is given in Ref.~\cite{Parrilo2003}.
For example, Eq.~\eqref{eq:sld-sos} provides an explicit matrix-SOS certificate. With the standard basis $\{\ket{e_a}\}_{a=1}^n$, it takes the form
\[
 \ket{f_{i,a}(\xi)}:=(\xi_iI-L_i^B)\rho^{1/2}\ket{e_a},\qquad
 Q_{B,K_B}(\xi)=\sum_{i=1}^d\sum_{a=1}^n
 \ket{f_{i,a}(\xi)}\bra{f_{i,a}(\xi)}
\]

For a fixed degree bound on the factors, the SOS condition can be expressed by positive semidefiniteness of a finite matrix.
For a nonnegative integer $k$ and a multi-index $\alpha=(\alpha_1,\ldots,\alpha_d)\in\mathbb N_0^d$, write
$|\alpha|:=\sum_i\alpha_i$ and $\xi^\alpha:=\prod_i\xi_i^{\alpha_i}$.
Throughout, the degree of a polynomial, polynomial vector, or matrix polynomial means its total degree: for $H(\xi)=\sum_\alpha H_\alpha\xi^\alpha$ with $H\not\equiv0$, set
\[
 \deg H:=\max_{\alpha:H_\alpha\ne0}|\alpha|
\]
and assign degree $-\infty$ to the zero polynomial. The polynomial $H$ is homogeneous of degree $k$ if $H_\alpha=0$ whenever $|\alpha|\ne k$, or equivalently if $H(\lambda\xi)=\lambda^kH(\xi)$ for every $\lambda\in\R\setminus\{0\}$.
Order the multi-indices $\alpha^{(1)},\alpha^{(2)},\ldots$ by increasing total degree, with a fixed order within each degree. The first $d+1$ monomials are ordered as $1,\xi_1,\ldots,\xi_d$.
Define the row vector of monomials of degree at most $k$ by
\[
 \nu_k:=\binom{d+k}{k},\qquad
 u_k(\xi):=\bigl(\xi^{\alpha^{(1)}},\ldots,\xi^{\alpha^{(\nu_k)}}\bigr)
 \in\R[\xi]^{1\times\nu_k}
\]
and its tensor product with the identity by
\[
 \mathcal U_k(\xi):=u_k(\xi)\otimes I_n
 =\begin{pmatrix}\xi^{\alpha^{(1)}}I_n&\cdots&\xi^{\alpha^{(\nu_k)}}I_n\end{pmatrix}
 \in\R[\xi]^{n\times n\nu_k}
\]
Every factor whose components have total degree at most $k$ can be written as $\ket{f_\nu(\xi)}=\mathcal U_k(\xi)c_\nu$, where $c_\nu\in\C^{n\nu_k}$ stacks its monomial coefficient vectors in this order.
A matrix-SOS representation with this degree bound is therefore equivalent to the existence of a constant matrix $G$ satisfying
\begin{equation}\label{eq:matrix-sos-gram}
 H(\xi)=\mathcal U_k(\xi)G\mathcal U_k(\xi)^{\dagger},\qquad G=G^{\dagger}\ge0
\end{equation}
The matrix $G$ is called a \emph{Gram matrix}; its dimension equals the number of columns of $\mathcal U_k$.
One direction follows by taking $G=\sum_\nu c_\nu c_\nu^{\dagger}$. Conversely, a factorization of $G$ gives an SOS representation by polynomial vectors.
Matching the coefficients of every monomial in this identity gives finitely many linear equations in $G$.
If the coefficients of $H$ depend affinely on unknown variables, these equations are real-linear in those variables and $G$; complex coefficients are matched separately in their real and imaginary parts.
Thus, with a linear objective, optimization subject to the SOS condition is an SDP expressed by $G\ge0$ and these coefficient equations~\cite{SH}. For the scalar case, see Ref.~\cite{Parrilo2003}.

We apply this construction to the pointwise positivity constraint in Eq.~\eqref{eq:Q}.
Set $w(\xi):=1+\|\xi\|^2$ and, for fixed $B$, define
\begin{equation}\label{eq:denom}
 d_m(B)=\inf_{X=X^{\dagger}}\{\tr X:w(\xi)^mQ_{B,X}(\xi)\text{ is a matrix SOS}\},\qquad m=0,1,\ldots
\end{equation}
This choice of multiplier is motivated by SOS representations with denominators for positive definite homogenizations; see Putinar and Vasilescu~\cite{PV} for the scalar case and Dinh, Ho, and Le~\cite{DHL} for the matrix case.
For $H(\xi)=\sum_{|\alpha|\le k}H_\alpha\xi^\alpha$ of degree at most $k$, its degree-$k$ homogenization is
\[
 \sum_{|\alpha|\le k}H_\alpha t^{k-|\alpha|}y^\alpha
 =t^kH(y/t)\quad(t\ne0),\qquad (t,y)\in\R\times\R^d
\]
where the left-hand side also defines it at $t=0$.
A feasible SOS representation yields
\[
 w(\xi)^mQ_{B,X}(\xi)=\sum_\nu\ket{f_\nu(\xi)}\bra{f_\nu(\xi)},\qquad
 Q_{B,X}(\xi)=\frac{\sum_\nu\ket{f_\nu(\xi)}\bra{f_\nu(\xi)}}{w(\xi)^m}
\]
We call this a \emph{matrix-SOS representation with a denominator}. Since $w>0$, every feasible point guarantees $Q_{B,X}(\xi)\ge0$ for all $\xi$.
The coefficients of $w^mQ_{B,X}$ depend affinely on $X$, so the Gram representation~\eqref{eq:matrix-sos-gram} makes each level a finite SDP. The factor degrees and the reduction of the Gram matrix according to the support and kernel are established in the next subsection.
The initial level $m=0$ uses sums of outer products of polynomial vectors of degree at most one. For the unreduced basis $\mathcal U_1(\xi)=(I_n,\xi_1I_n,\ldots,\xi_dI_n)$, write the Gram matrix as $G=(G_{ij})_{i,j=0}^d$, with $G_{ij}\in\C^{n\times n}$. Its constraints are
\begin{equation}\label{eq:initial-gram}
 \begin{gathered}
 G\ge0,\quad G_{00}=X,\quad G_{ii}=\rho\quad(1\le i\le d),\\
 G_{0i}+G_{i0}=-2A_i\quad(1\le i\le d),\qquad
 G_{ij}+G_{ji}=0\quad(1\le i<j\le d)
 \end{gathered}
\end{equation}
Equation~\eqref{eq:sld-sos} provides a feasible point.
Increasing only the allowed factor degree for $Q_X$ itself does not strengthen the initial level: the sum of squared moduli of the highest-degree terms cannot cancel, so the factors must still have degree at most one.
The positive multipliers $w^m$ enlarge the class of certificates.

\begin{proposition}\label{prop:hierarchy-monotonicity}
We have $h_B\le d_{m+1}(B)\le d_m(B)<\infty$, and
$c_m(W;B)=2\Tr(BW^{1/2})-d_m(B)$ is a monotonically nondecreasing lower bound on the risk.
\end{proposition}
\begin{proof}
If $w^mQ_X=\sum_\nu\ket{f_\nu}\bra{f_\nu}$, then
$w^{m+1}Q_X=\sum_\nu\ket{f_\nu}\bra{f_\nu}+\sum_{i,\nu}\xi_i^2\ket{f_\nu}\bra{f_\nu}$ is also an SOS.
Thus the feasible sets are nested and increasing. The bound $h_B\le d_m(B)$ follows from Theorem~\ref{thm:exact-dual}, and finiteness follows from the SLD certificate at the initial level.
\end{proof}

\subsection{Effective subspaces and degree bounds on the support and kernel}\label{sec:effective-space}
For fixed $B$, we may compress the problem to $\mathcal K_B:=P\mathcal H+\sum_i\ran(QA_iP)$.
Let $\Pi_B$ be the orthogonal projection onto $\mathcal K_B$, and define the compression of an operator $T$ as $(\Pi_BT\Pi_B)|_{\mathcal K_B}$.
Since
\[
 \Pi_B\rho\Pi_B=\rho,\qquad \Pi_BA_i\Pi_B=A_i
\]
compressing a measurement leaves $\tr(\rho M_x)$ and $\tr(A_iM_x)$ unchanged.
Conversely, extending a measurement on $\mathcal K_B$ by zero and adjoining the effect $I-\Pi_B$ gives a POVM on the original space. The additional effect has zero probability and zero trace against each $A_i$. Hence $h_B$ is unchanged.
Projecting and extending the SOS factors by zero likewise preserve $d_m(B)$. Under projection, the trace of $X$ does not increase because $X\ge0$.
When $B$ varies, we use the common subspace obtained by replacing $A_i$ with $D_j$.
In the degree bounds and convergence arguments below, we use the same notation for the compressed operators and write $n$ for the dimension of the effective subspace.

Set $r:=\rank\rho$ and $\ell:=n-r$, and decompose the operators according to the support and kernel:
\begin{equation}\label{eq:blocks}
 \rho=\begin{pmatrix}\rho_0&0\\0&0\end{pmatrix},\quad
 A_i=\begin{pmatrix}A_{0i}&A_{1i}\\A_{1i}^{\dagger}&0\end{pmatrix},\quad
 X=\begin{pmatrix}X_0&X_1\\X_1^{\dagger}&X_2\end{pmatrix},\quad \rho_0>0,\quad n=r+\ell
\end{equation}
Here $\rho_0$ is the positive definite block on the support.
The matrices $\rho_0,A_{0i},X_0$ are $r\times r$, $A_{1i},X_1$ are $r\times\ell$, and $X_2$ is $\ell\times\ell$. The matrices $X_0,X_1,X_2$ denote the support, cross, and kernel blocks, respectively.
Also set
\[
 A_a(\xi):=\sum_{i=1}^d\xi_iA_{ai}\qquad(a=0,1)
\]
We do not compress to the support alone, since that would discard $A_{1i}$.
Using the monomial row vector $u_k$, define a basis with separate degree bounds on the support and kernel by
\begin{equation}\label{eq:graded-basis}
 \begin{split}
 V_m(\xi)&:=\begin{pmatrix}
 u_{m+1}(\xi)\otimes I_r&0\\
 0&u_m(\xi)\otimes I_\ell
 \end{pmatrix},\qquad N_m:=r\nu_{m+1}+\ell\nu_m,\\
 \mathcal I_m&:=\{(a,\alpha):1\le a\le r,\ |\alpha|\le m+1\}
 \cup\{(a,\alpha):r<a\le n,\ |\alpha|\le m\}
 \end{split}
\end{equation}
The matrix $V_m$ has size $n\times N_m$; the kernel block is omitted when $\ell=0$.
Its columns are indexed by $\mathcal I_m$. The support block precedes the kernel block. Within each block, the columns follow the monomial order in $u_k$, and then the component index $a$ for each monomial. We use the same ordering for coefficient vectors and Gram-matrix indices.

The aim is to reduce the Gram-matrix size while preserving the SOS condition and ensuring the existence of a positive definite feasible Gram matrix.
The basis $\mathcal U_k$ in Sec.~\ref{sec:sos-definition} gives a Gram matrix of size $n\nu_k\times n\nu_k$, so unnecessary degrees increase computational and memory costs.
The next lemma shows that all coefficients above degree $m+1$ in the support components and degree $m$ in the kernel components must vanish, whereas the remaining basis $V_m$ admits a positive definite Gram matrix.

\Needspace{8\baselineskip}
\begin{lemma}[Exact reduction of the Gram representation and strictly feasible Gram matrices]\label{lem:graded-degree}\label{lem:strict-gram}
For fixed $B$ and $m\in\mathbb N_0$, the following statements hold.
\begin{enumerate}
\item[(i)] For any Hermitian matrix $X$, every finite SOS representation
\[
 w^mQ_{B,X}=\sum_{\nu=1}^N\ket{f_\nu}\bra{f_\nu},\qquad
 \ket{f_\nu(\xi)}=\sum_{a=1}^n\sum_{\alpha\in\mathbb N_0^d}c_{\nu,a,\alpha}\xi^\alpha\ket{e_a}
\]
of $w^mQ_{B,X}$ satisfies
\[
 c_{\nu,a,\alpha}=0\qquad(1\le\nu\le N,\ (a,\alpha)\notin\mathcal I_m)
\]
Here $\{\ket{e_a}\}_{a=1}^n$ is the standard basis adapted to the support--kernel decomposition in Eq.~\eqref{eq:blocks}.
Consequently,
\[
 \begin{gathered}
 w^mQ_{B,X}\text{ is a matrix SOS}
 \quad\Longleftrightarrow\quad
 \exists G=G^{\dagger}\ge0:\\
 w^mQ_{B,X}=V_mGV_m^{\dagger},\qquad G\in\C^{N_m\times N_m}.
 \end{gathered}
\]
\item[(ii)] There exist a Hermitian matrix $X^{(m)}\in\C^{n\times n}$ and a constant matrix $G_m\in\C^{N_m\times N_m}$ such that
\[
 w^mQ_{B,X^{(m)}}=V_mG_mV_m^{\dagger},\qquad G_m=G_m^{\dagger}>0
\]
\end{enumerate}
\end{lemma}
The proof is given in Appendix~\ref{app:hierarchy-degree-proof}. The degree bounds use the absence of cancellation in sums of squared norms. Strict feasibility is obtained from a positive definite Gram matrix at $m=0$, followed by coordinate embeddings associated with the scalar SOS multiplier $w^m$.

By Lemma~\ref{lem:graded-degree}(i), the common degree bound $k=m+1$ is sufficient. Using $V_m$ further reduces the Gram-matrix size from $n\nu_{m+1}\times n\nu_{m+1}$ to $N_m\times N_m$.
This reduction preserves the SOS condition and the optimal value $d_m(B)$, so Eq.~\eqref{eq:denom} is equivalent to the finite SDP
\begin{equation}\label{eq:gram-sdp}
 \inf_{X,G}\{\tr X:w^mQ_{B,X}=V_mGV_m^{\dagger},\quad X=X^{\dagger},\quad G\ge0\}
\end{equation}
The identity is imposed by matching all monomial coefficients, without sampling points.
If $P_{\gamma,ab}$ denotes the coefficient of $\xi^\gamma$ in the $(a,b)$ entry on the left, then
$P_{\gamma,ab}=\sum_{\alpha+\beta=\gamma}G_{(a,\alpha),(b,\beta)}$, where the sum is restricted to $(a,\alpha),(b,\beta)\in\mathcal I_m$.
Lemma~\ref{lem:graded-degree}(ii) establishes \emph{Slater's condition} for this SDP: there is a feasible point satisfying the coefficient equations with $G>0$.
Together with the finiteness of the optimal value from Proposition~\ref{prop:hierarchy-monotonicity}, this implies equality of the primal and dual values and attainment of the dual optimum~\cite{BoydVandenberghe}.

For a feasible Gram matrix $\widehat G$ in the larger basis $\mathcal U_k$, with $k\ge m+1$, a factorization gives coefficients as in part (i), and hence
\[
 \widehat G_{(a,\alpha),(a,\alpha)}
 =\sum_{\nu=1}^N|c_{\nu,a,\alpha}|^2=0
 \qquad((a,\alpha)\notin\mathcal I_m,\ |\alpha|\le k)
\]
Thus, when $\rho$ is singular, a Gram matrix in $\mathcal U_{m+1}$ cannot be positive definite. Even when $\rho>0$, positive definiteness is impossible if $k>m+1$.
Adding excess degrees at a fixed $m$ does not enlarge the feasible set of $X$. The hierarchy instead increases the exponent $m$ of the multiplier $w^m$.

\subsection{Convergence with a singular highest-degree coefficient}\label{sec:convergence}
The next lemma extends the SOS theorem with denominators for real symmetric matrices of Dinh, Ho, and Le~\cite{DHL} to complex Hermitian matrices by realification.
\begin{lemma}[Matrix-SOS representation with a denominator; complex Hermitian version~{\cite{DHL}}]\label{lem:matrix-denominator}
Let $p,q\in\mathbb N$, $k\in\mathbb N_0$, and let $F\in\C[z_1,\ldots,z_q]^{p\times p}$ be a homogeneous Hermitian matrix polynomial of degree $2k$.
If $F(z)>0$ for every $z\in\R^q\setminus\{0\}$, there exist $m\in\mathbb N_0$, $N\in\mathbb N$, and polynomial vectors $\ket{f_\nu}\in\C[z_1,\ldots,z_q]^p$ such that
\[
 (z_1^2+\cdots+z_q^2)^mF(z)
 =\sum_{\nu=1}^N\ket{f_\nu(z)}\bra{f_\nu(z)}
\]
\end{lemma}
The realification argument is given in Appendix~\ref{app:matrix-denominator-proof}.

The conclusion does not hold in general under positive semidefiniteness alone.
Even in the scalar case, there exist nonnegative homogeneous polynomials $f$ for which $\|z\|^{2m}f(z)$ is not SOS for any $m\in\mathbb N_0$~\cite{ReznickNoUniform}.
Accordingly, the convergence proof in Appendix~\ref{app:hierarchy-convergence-proof} adds $\delta I_n$ with $\delta>0$ to an optimal solution $X_*$ of Theorem~\ref{thm:exact-dual}. Separating the support and kernel and taking a Schur complement then ensures positive definiteness of the homogeneous matrix polynomial to which the lemma is applied.

\begin{theorem}\label{thm:denom-convergence}
For a state of any rank, $d_m(B)\downarrow h_B$.
\end{theorem}
The proof is given in Appendix~\ref{app:hierarchy-convergence-proof}.
When $\rho$ is singular, the usual homogenization $t^2X+\|y\|^2\rho-2tA(y)$, with $t\in\R$ and $y\in\R^d$, vanishes in kernel directions at $t=0$.
Lemma~\ref{lem:matrix-denominator} therefore cannot be applied directly to this homogenization.
The proof first separates the kernel block, applies the lemma on the support, and then reconstructs the full matrix by completing the square.
Only $X$ is perturbed; $\rho,D_j$ remain unchanged. No finite-level exactness or uniform convergence rate is asserted.

\subsection{Blockwise SOS hierarchies using a sufficient algebra}\label{sec:sos-sufficiency}
If a sufficient algebra of the form~\eqref{eq:barc-sufficient-blocks} in Sec.~\ref{sec:barc-sufficiency} is known, the lower-bound computation also decomposes into SDPs on the reduced blocks.
The numerical experiments in Sec.~\ref{sec:sos-block-evidence} apply this decomposition to Schur blocks.
Using the unnormalized data $\rho_b,D_{j,b}\in\mathcal{B}_h(\C^{n_b})$ from Sec.~\ref{sec:barc-sufficiency} and a common $B\in\R^{d\times d}$, set
\begin{equation}\label{eq:block-sos-polynomial}
 A_{i,b}:=\sum_{j=1}^dB_{ij}D_{j,b},\qquad
 Q^{(b)}_{B,X_b}(\xi):=X_b+\|\xi\|^2\rho_b-2\sum_{i=1}^d\xi_iA_{i,b}
\end{equation}
Here $X_b\in\mathcal{B}_h(\C^{n_b})$. Keeping $\rho_b$ unnormalized retains the Fisher information arising from derivatives of the block probabilities.
Let $M_b$ be a POVM on block $b$, and write its Fisher information as $F_b[M_b]$ with the convention in Eq.~\eqref{eq:barc-block-fisher}. Then
\begin{equation}\label{eq:block-support-sum}
 h_B=\sum_{b=1}^q h_{B,b},\qquad
 h_{B,b}:=\sup_{M_b}\Tr\bigl(BF_b[M_b]B^{\mathsf T}\bigr).
\end{equation}
Indeed, sufficiency allows any measurement to be mapped into the algebra, and refinement that retains the block label as an outcome cannot decrease the Fisher information.
Conversely, measurements chosen independently in the blocks define a POVM on the original space, proving the equality.
If $\rho_b=0$, positivity of the original local model implies $D_{j,b}=0$ for every $j$. This block contributes zero and may be omitted.

In a basis adapted to the support and kernel of each block, set $r_b:=\rank\rho_b$ and $\ell_b:=n_b-r_b$, and use
\[
 V_{m,b}(\xi):=\diag\bigl(u_{m+1}(\xi)\otimes I_{r_b},\,
                              u_m(\xi)\otimes I_{\ell_b}\bigr),\qquad
 N_{m,b}:=r_b\nu_{m+1}+\ell_b\nu_m
\]
By Lemma~\ref{lem:graded-degree}(i), the reduced hierarchy is computed by the blockwise finite SDPs
\begin{equation}\label{eq:block-sos-hierarchy}
 \begin{split}
 d_m^{\mathcal A}(B)&:=\sum_{b=1}^q d_{m,b}(B),\\
 d_{m,b}(B)&:=\inf_{X_b,G_b}\left\{\tr X_b:
 w^mQ^{(b)}_{B,X_b}=V_{m,b}G_bV_{m,b}^{\dagger},\quad
 G_b=G_b^{\dagger}\ge0\right\}
 \end{split}
\end{equation}
The identities are imposed by coefficient matching within each block.
This hierarchy satisfies
\begin{equation}\label{eq:block-sos-convergence}
 h_B\le d_{m+1}^{\mathcal A}(B)\le d_m^{\mathcal A}(B)\le d_m(B),\qquad
 d_m^{\mathcal A}(B)\longrightarrow h_B
\end{equation}
The first inequality follows from completing the square in each block and Eq.~\eqref{eq:block-support-sum}. Monotonicity follows by multiplication by $w$, as in Proposition~\ref{prop:hierarchy-monotonicity}.
For the last inequality, take a feasible $X$ on the original space $\mathcal H$, extending it by zero if it has already been compressed to an effective subspace, and set
\[
 X_b:=\tr_{\C^{m_b}}(\iota_b^{\dagger}X\iota_b),\qquad
 w^mQ^{(b)}_{B,X_b}
 =\tr_{\C^{m_b}}\bigl(\iota_b^{\dagger}(w^mQ_{B,X})\iota_b\bigr)
\]
Compression and partial trace preserve matrix-SOS representations, and $\sum_b\tr X_b=\tr X$.
Convergence then follows from Theorem~\ref{thm:denom-convergence}, and $2\Tr(BW^{1/2})-d_m^{\mathcal A}(B)$ is a lower bound on the risk in the original estimation problem.
For fixed $B$, the SDPs can be solved independently and in parallel.
If $B$ is also a variable, share the same $B$ across all blocks and maximize
$2\Tr(BW^{1/2})-\sum_b\tr X_b$ subject to Eq.~\eqref{eq:block-sos-hierarchy}. This is a single SDP.
Convergence to the optimal risk in this case is discussed in Sec.~\ref{sec:joint}.

This construction does not retain the multiplicity spaces of dimensions $m_b$, and its semidefinite constraints involve $q$ Gram matrices of dimensions $N_{m,b}$.
Compared with using a single dense Gram matrix on the entire reduced space $\bigoplus_b\C^{n_b}$, the number of real variables and the memory requirement decrease from
$\bigl(\sum_bN_{m,b}\bigr)^2$ to $\sum_bN_{m,b}^2$.
Matrix operations such as eigendecompositions can also be performed blockwise, reducing the computational cost of the SDP.
Compression to the effective subspace described in Sec.~\ref{sec:effective-space} can additionally be applied within each block.
\section{Joint optimization of the coefficients: from NH to the optimal risk}\label{sec:joint}
The Nagaoka--Hayashi (NH) bound $c_{\mathrm{NH}}(W)$ is the lower bound on the weighted risk of locally unbiased estimation defined by Eq.~\eqref{eq:nh-primal} in Appendix~\ref{app:nh}.
\subsection{A single SDP containing the fixed-coefficient formulations}
The polynomial $Q_{B,X}$ is affine in $B,X$. Fix a basis in the common subspace
$\mathcal K=P\mathcal H+\sum_j\ran(QD_jP)$.
In this section, $\rho,D_j$ denote the operators compressed to this subspace, and
$V_m,N_m$ are defined from its support--kernel decomposition as in Eq.~\eqref{eq:graded-basis}.
For variables $B\in\R^{d\times d}$, $X\in\mathcal{B}_h(\mathcal K)$, and $G\in\mathcal{B}_h(\C^{N_m})$, define
\begin{equation}\label{eq:joint-sdp}
 c_m(W)=\sup_{B,X,G}\left\{2\Tr(BW^{1/2})-\tr X:
 w^mQ_{B,X}=V_mGV_m^{\dagger},\quad G\ge0\right\}
\end{equation}
There are no products of unknown variables, so this is a single SDP.
It is not a nonconvex formulation treating $B^{\mathsf T}B$ as an independent variable.
For any fixed $B_0\in\R^{d\times d}$,
\[
 c_m(W)=\sup_B c_m(W;B)\ge c_m(W;B_0)
\]
Thus the joint formulation gives a bound at least as strong as any fixed-$B$ formulation at the same level. This comparison concerns exact optimal values; the ordering is not guaranteed for numerical bounds obtained after finite-precision solution and certificate correction.

\subsection{Initial-level equality with the weighted NH bound}
\begin{theorem}\label{thm:joint-convergence}
For a quantum statistical manifold as introduced in Sec.~\ref{sec:model},
\begin{equation}\label{eq:joint-chain}
 c_0(W)=c_{\mathrm{NH}}(W)\le c_1(W)\le c_2(W)\le\cdots\le c_{\theta_0}(W),
 \qquad\lim_{m\to\infty}c_m(W)=c_{\theta_0}(W).
\end{equation}
\end{theorem}
\begin{proof}
Equality between the initial level and the NH bound follows from Theorem~\ref{thm:nh-dual-sos}.
Weak duality at each level follows from Eq.~\eqref{eq:square}, and monotonicity follows from inclusion of the SOS cones.
To prove convergence, let $\mathcal F$ be the closure of the set of Fisher information matrices of finite-outcome measurements.
By Proposition~\ref{prop:sld-randomization}, $\mathcal F$ contains a positive definite matrix.
Randomizing measurements while retaining the measurement choice as part of the outcome produces convex combinations of their Fisher matrices, so $\mathcal F$ is convex.
Monotonicity of the SLD metric, Eq.~\eqref{eq:sld-measurement}, gives $0\le F\le J_{\theta_0}$; hence the closure $\mathcal F$ is compact.
Since $W>0$, $\Tr(WF^{-1})\to+\infty$ as $F$ approaches a singular matrix.
There is therefore a positive definite minimizer $F_*$ on $\mathcal F$, and continuity of inversion implies that its value is $c_{\theta_0}(W)$.

The one-sided directional derivative toward any $F\in\mathcal F$ is nonnegative, so
\[
 0\le-\Tr\{F_*^{-1}WF_*^{-1}(F-F_*)\}.
\]
Setting $B_*=W^{1/2}F_*^{-1}$ gives
\[
 h_{B_*}=\Tr(B_*F_*B_*^{\mathsf T})=c_{\theta_0}(W),\qquad
 \Tr(B_*W^{1/2})=c_{\theta_0}(W).
\]
Combining these equalities with weak duality from Eq.~\eqref{eq:square} yields the exact variational representation
\begin{equation}\label{eq:exact-variational}
 c_{\theta_0}(W)=\sup_{B\in\R^{d\times d}}
 \{2\Tr(BW^{1/2})-h_B\}
\end{equation}
Applying Theorem~\ref{thm:denom-convergence} to this single $B_*$ and using
\[
 2\Tr(B_*W^{1/2})-d_m(B_*)\le c_m(W)\le c_{\theta_0}(W)
\]
establishes convergence. The argument does not require $F_*$ to be attained by an actual POVM.
\end{proof}

Monotonicity of the hierarchy does not imply strict improvement at every level. In models where the NH bound is attainable, every level equals the NH bound.
For fixed $B$, the limiting bound is $2\Tr(BW^{1/2})-h_B$, which need not equal $c_{\theta_0}(W)$.
Joint optimization of the coefficients closes this gap in the limit, but equality at a finite level is not guaranteed.

\subsection{Certification and computational size}
On the same common subspace, the joint formulation adds only the $d^2$ real variables in $B$ to the fixed-coefficient formulation; the Gram dimension $N_m$ is unchanged.
The fixed-coefficient formulation may, however, admit a further reduction to $\mathcal K_B$.
When a sufficient algebra is available, the blockwise SDPs in Sec.~\ref{sec:sos-sufficiency} give
\[
 \begin{gathered}
 c_m^{\mathcal A}(W):=\sup_B\left\{2\Tr(BW^{1/2})-d_m^{\mathcal A}(B)\right\},\\
 c_m(W)\le c_m^{\mathcal A}(W)\le c_{\theta_0}(W),\qquad
 c_m^{\mathcal A}(W)\uparrow c_{\theta_0}(W).
 \end{gathered}
\]
The comparison follows from Eq.~\eqref{eq:block-sos-convergence}, and convergence follows from Theorem~\ref{thm:joint-convergence}.
All blocks share the same $B$, and the objective is $2\Tr(BW^{1/2})-\sum_b\tr X_b$; this differs from choosing $B$ independently in each block.

Any point satisfying the SOS constraints yields a lower bound, whether or not it is optimal.
In floating-point arithmetic, errors may affect both positive semidefiniteness of the Gram matrix and the coefficient equations, so a solver's termination status alone does not constitute a mathematical certificate.
If a certified upper bound $\overline h\ge h_B$ on the support function is available, downward evaluation of $2\Tr(BW^{1/2})-\overline h$ gives a certified lower bound by Eq.~\eqref{eq:fixed-bound}.
For the joint formulation as well, certification fixes the computed $B$ and accounts for evaluation errors in $A_i=\sum_jB_{ij}D_j$.
Denote the resulting certified lower bound by $\underline c_{\mathrm{cert}}$.
If, in addition, $\overline c_{\mathrm{cert}}$ is a risk upper bound obtained by certifying POVM completeness and positivity and the Fisher computation, then
\[
 \underline c_{\mathrm{cert}}\le c_{\theta_0}(W)\le \overline c_{\mathrm{cert}}
\]
Numerically corrected lower bounds need not be monotone in the level, so we retain the maximum of the certified lower bounds obtained from all computations, including those with fixed coefficients.
\section{Numerical experiments}\label{sec:demo}
After stating the common conventions and certification scope, we report Experiments 1--4 for qubit and qutrit models, benchmark the two-stage measurement optimizer, and summarize the computation times.

\subsection{Common conventions and scope of certification in Experiments 1--4}\label{sec:experiment-conventions}
In each experiment, we construct the locally unbiased estimator in Eq.~\eqref{eq:efficient-estimator} from the Fisher information at the evaluation point and compare its weighted covariance cost.
We write $c_{\mathrm{num}}:=\Tr(WF_{\theta_0}[\widehat M]^{-1})$ for the floating-point evaluation of a numerical candidate measurement $\widehat M$,
$\underline c_1$ for a certified risk lower bound obtained from level $m=1$ and valid for all POVMs, and $\overline c_{\mathrm{NH}}$ for a certified upper bound on the NH value obtained from a feasible point of the NH primal problem.
The certified relations are
\begin{equation}\label{eq:experiment-certification}
 \begin{gathered}
 \underline c_1\le c_{\theta_0}(W),\qquad c_{\mathrm{NH}}(W)\le \overline c_{\mathrm{NH}},\\
 \underline c_1>\overline c_{\mathrm{NH}}\ \Longrightarrow\ c_{\theta_0}(W)-c_{\mathrm{NH}}(W)\ge \underline c_1-\overline c_{\mathrm{NH}}>0.
 \end{gathered}
\end{equation}
Thus $\overline c_{\mathrm{NH}}$ is an upper bound on the value of the NH relaxation, not an upper bound on the risk of an attainable measurement.
Because $c_{\mathrm{num}}$ is not accompanied by a rigorous upper certificate, $c_{\mathrm{num}}-\underline c_1$ measures the separation from the numerical candidate and is not a certified upper bound on the optimality gap.

In Experiments 2--4, $\underline c_1$ is obtained from numerical solutions at level $m=1$ of the denominator-based matrix-SOS hierarchy.
We treat the Gram factors as dyadic rationals, compute their products exactly, and bound the coefficient residuals to correct the result into a lower bound valid for all POVMs.
We further correct errors between the numerical inputs and the ideal states, derivatives, and weights using rational arithmetic and outward-rounded interval arithmetic.
On the NH side, $\overline c_{\mathrm{NH}}$ is obtained from a primal feasible point for the same ideal model by repairing local unbiasedness and positive semidefiniteness.
This procedure certifies Eq.~\eqref{eq:experiment-certification}; it does not assert that the corrected values are the values of exactly feasible points in the original finite SOS cone or exact SDP optima.
In the tables, $\underline c_1$ is rounded downward and $\overline c_{\mathrm{NH}}$ upward to 12 decimal places, whereas differences are computed from the rational values before rounding.

Measurement optimization used 10 initializations (seeds $0,\ldots,9$) for each point or example, and we selected the smallest final cost among the runs.
Adam used 30000 updates, and the RCG gradient stopping threshold was $10^{-9}$. Settings that differ between experiments are listed in Table~\ref{tab:experiment-settings}.
General weights were reduced to unit weight by the coordinate transformation in Sec.~\ref{sec:barc-weighted-implementation}.
The number $s$ is the row count of the Stiefel matrix used in the optimization and counts outcomes before removing small effects or merging proportional effects. It is not the minimum number of outcomes required.
Termination at the iteration limit or by a failed line search is distinguished from satisfaction of the gradient stopping criterion.
\begin{table}[tbp]
\centering\small
\caption{Measurement-optimization settings in Experiments 1--4. All cases use 10 initializations and 30000 Adam updates.}\label{tab:experiment-settings}
\begin{tabular}{@{}lrrr@{}}
\toprule
Experiment & Outcomes $s$ & Adam rate & RCG limit\\
\midrule
1 and 2 & $2^{N+1}$ & $0.005$ & $2000$\\
3: diagonal weight & $24$ & $0.005$ & $8000$\\
3: nondiagonal weight & $32$ & --- & ---\\
4: examples A and B & $48$ & --- & ---\\
\bottomrule
\end{tabular}
\par\smallskip\footnotesize
--- denotes a run-time setting that could not be identified.
\end{table}

OpenAI Codex (Astra 6) was used under the authors' direction to assist with manuscript preparation, refinement of mathematical derivations, code development, and the execution and analysis of numerical experiments.
The authors verified the resulting content, code, and numerical results and take full responsibility for this work.

\subsection{Experiment 1: Qubit measurements with SLD Fisher weighting}\label{sec:experiment1}
We consider the full three-parameter qubit model and its $N$-copy extension,
\begin{equation}\label{eq:experiment-qubit-model}
 \rho_\theta^{(1)}=\frac{I+\theta^1\sigma_x+\theta^2\sigma_y+\theta^3\sigma_z}{2},\qquad
 \rho_\theta^{(N)}=(\rho_\theta^{(1)})^{\otimes N},\qquad
 \theta_0=(0,0,r).
\end{equation}
We use the usual Pauli matrices, with $\sigma_y=\left(\begin{smallmatrix}0&-i\\i&0\end{smallmatrix}\right)$ in the basis $\{\ket0,\ket1\}$.
We examined 50 points: $N=2,3,4,5,6$ and $r=0,0.1,\ldots,0.9$.
In standard Bloch coordinates, the SLD Fisher information is
\[
 J_{\theta_0}^{(1)}=\diag\bigl(1,1,(1-r^2)^{-1}\bigr),\qquad
 J_{\theta_0}^{(N)}=N J_{\theta_0}^{(1)},
\]
and we take the weight to be $W=J_{\theta_0}^{(N)}$.
Let $V_{\theta_0}^{(N)}$ denote the covariance obtained by combining the measurement $M$ with the locally unbiased estimator in Eq.~\eqref{eq:efficient-estimator},
and abbreviate $F:=F_{\theta_0}[M]$. For this estimator, $V_{\theta_0}^{(N)}=F^{-1}$.
Set $T:=\diag(1,1,\sqrt{1-r^2})$ and introduce local coordinates by $\theta=\theta_0+T\eta$, with $\eta\in\R^3$.
Writing the Fisher information in these coordinates as $\widetilde F=TFT$, we compare
\begin{equation}\label{eq:experiment-sld-risk}
 \Tr\bigl(J_{\theta_0}^{(N)}V_{\theta_0}^{(N)}\bigr)
   =\Tr\bigl(J_{\theta_0}^{(N)}F^{-1}\bigr)
   =N\Tr\widetilde F^{-1}.
\end{equation}
This is the weighted variance per batch of $N$ copies.
We minimize $\Tr\widetilde F^{-1}$ and report $N$ times its floating-point evaluation as $c_{\mathrm{num}}$.

We compare the numerical Q-StARC candidate, a numerical solution for the NH bound, and the finite Vidal POVM\cite{VidalMixed,LPT}.
To specify the finite Vidal measurement, write the Schur--Weyl decomposition as
\[
 \begin{aligned}
 (\C^2)^{\otimes N}
 &\simeq\bigoplus_{q=N,N-2,\ldots,N\bmod2}
 \operatorname{Sym}^{q}(\C^2)\otimes\C^{\mu_{N,q}},\\
 \mu_{N,q}&:=\binom N{(N-q)/2}-\binom N{(N-q)/2-1},
 \end{aligned}
\]
with binomial coefficients outside their allowed range set to zero.
Here $\operatorname{Sym}^{q}(\C^2)$ is the symmetric subspace of $q$ qubits, a $(q+1)$-dimensional spin-$j=q/2$ space,
and $\mu_{N,q}$ is its multiplicity.
For a unit direction $n=(n_1,n_2,n_3)\in\R^3$, define the spin-coherent state by
\[
 \ket n\bra n=\frac{I+n_1\sigma_x+n_2\sigma_y+n_3\sigma_z}{2},\qquad
 \ket{q/2,n}:=\ket n^{\otimes q}\in\operatorname{Sym}^{q}(\C^2)
 \quad(q\ge1).
\]
For $q=0$, the space is $\C$, the state is $1$, and no direction is assigned.
Using the directions $n_{q,x}$ and positive weights $c_{q,x}^2$ listed in Appendix~\ref{app:finite-vidal}, the effects and their probabilities are
\begin{equation}\label{eq:finite-vidal-probability}
 \begin{split}
 M_{q,x}&=c_{q,x}^2\ket{j,n_{q,x}}\bra{j,n_{q,x}}\otimes I_{\mu_{N,q}},\\
 p_\theta(q,x)&=c_{q,x}^2\mu_{N,q}
 \left(\frac{1-\|\theta\|^2}{4}\right)^{(N-q)/2}
 \left(\frac{1+\theta\cdot n_{q,x}}2\right)^q,
 \end{split}
\end{equation}
as in Ref.~\cite{VidalMixed}. For $q=0$, the last factor in the probability formula is set to $1$.
Each effect is embedded in its corresponding Schur block and is zero on all other blocks.
For $q=2,\ldots,6$, we use the directions and weights in Ref.~\cite{LPT} in their original orientations. For $q=1$, we use the two directions $\pm(1,1,1)/\sqrt3$; for $q=0$, we use a single effect. All weights for $q=0,1$ are $c_{q,x}^2=1$.
The directions are fixed independently of $N$ and $r$, with no randomization over rotations.
The outcome counts for $N=2,\ldots,6$ are $5,8,15,20,33$, respectively; multiplicity labels are not treated as separate outcomes.
We combine this measurement with the locally unbiased estimator in Eq.~\eqref{eq:efficient-estimator} and evaluate $F^{-1}$. The estimator optimized for Bayes fidelity in the original paper is not used in this comparison.

\begin{figure}[htbp]
\centering
\includegraphics[width=\linewidth]{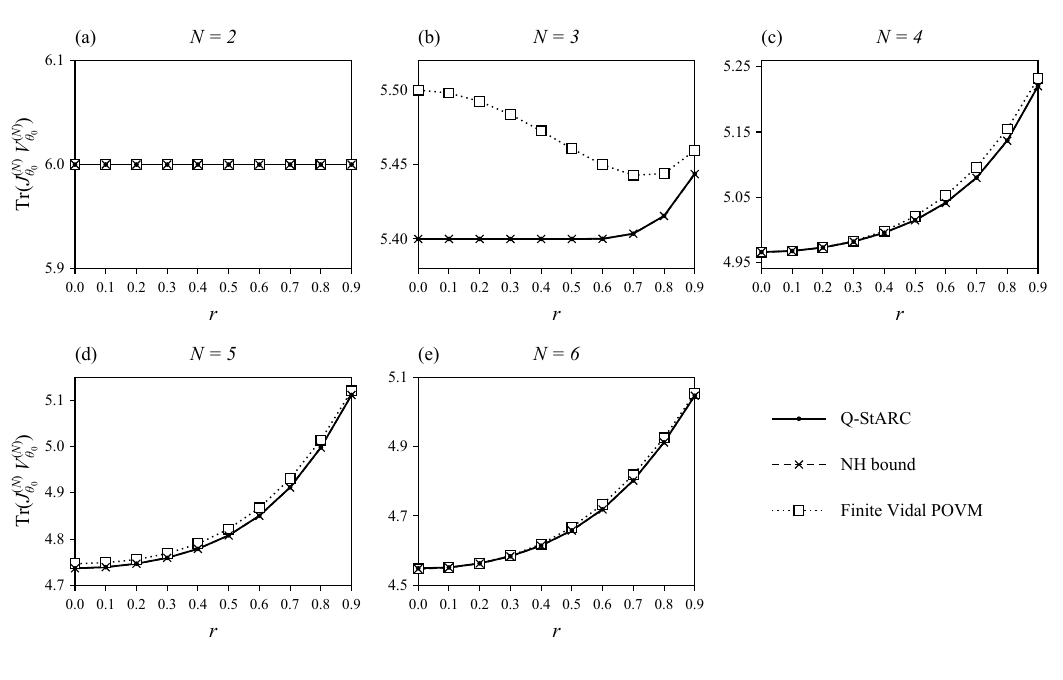}
\caption{Experiment 1: Weighted variance $\Tr(J_{\theta_0}^{(N)}V_{\theta_0}^{(N)})$ with the $N$-copy SLD Fisher weight.
Q-StARC denotes the best numerical candidate from 10 initializations at each point; NH denotes the numerical relaxation value; finite Vidal denotes the fixed-direction POVM combined with a locally unbiased estimator.
The three curves nearly coincide for $N=2$. The vertical-axis range differs between panels, and the lines connect the 10 computed points in each panel.}
\label{fig:experiment1}
\end{figure}

Figure~\ref{fig:experiment1} shows the results at all 50 points.
For example, at $N=3,r=1/2$, the Q-StARC candidate cost is $5.400000000000$, compared with $5.460674157303$ for the finite Vidal measurement.
For $N=6$ at $r=0,0.1$, however, the Q-StARC candidate costs are $4.549199820861,\allowbreak4.551606675973$, whereas the finite Vidal costs are $4.547368421053,\allowbreak4.551241299894$, respectively, and are lower.
Thus the Q-StARC candidates do not improve on the finite Vidal measurement at every evaluation point.
Among the 50 selected runs, 31 stopped at the iteration limit and 19 stopped because the line search failed.
The NH solver reported \texttt{optimal\_inaccurate} at every point, and the NH values in this figure are not rigorously certified.
We do not infer strict separation or global optimality from the small differences in the figure; the next experiment provides a certified comparison at $r=1/2$.

\subsection{Experiment 2: Unattainability of the NH bound with SLD Fisher weighting}\label{sec:experiment2}\label{sec:nh-evidence}\label{sec:sos-block-evidence}
We set $r=1/2$ in Eq.~\eqref{eq:experiment-qubit-model} and use the same selected measurements as in Experiment 1 for $N=2,\ldots,6$.
We fix $B=\widehat{\widetilde F}^{-1}$ in SLD-normalized coordinates and solve the $m=1$ SOS problem in each Schur block of the sufficient subalgebra described in Sec.~\ref{sec:sos-sufficiency}.
The coefficient $B$ is common to all blocks and is not jointly optimized in this experiment.
The lower bound obtained for the internal unit-weight problem is multiplied by $N$ to match the scale of Eq.~\eqref{eq:experiment-sld-risk}.
Measurement optimization uses a dense Stiefel representation in dimension $2^N$; Schur reduction is applied on the SOS side.

The certified values in Table~\ref{tab:experiment2-risk} include corrections for input errors relative to the ideal model at $r=1/2$.
For $N=4,5,6$, we certify
\begin{equation}\label{eq:experiment2-nh-gap}
 c_{\theta_0}(J_{\theta_0}^{(N)})-c_{\mathrm{NH}}(J_{\theta_0}^{(N)})
 \ge
 \begin{cases}
 0.000148589,&N=4,\\
 0.000121917,&N=5,\\
 0.000387471,&N=6
 \end{cases}.
\end{equation}
This establishes that the NH bound is unattainable for full three-parameter estimation in these cases.
For $N=2,3$, the certificates in this experiment do not establish $\underline c_1>\overline c_{\mathrm{NH}}$, nor do they establish equality.
For $N=2,\ldots,6$, the differences $c_{\mathrm{num}}-\underline c_1$ are approximately
$1.49\times10^{-6},6.90\times10^{-7},2.76\times10^{-7},1.25\times10^{-6},2.17\times10^{-6}$, respectively.
\begin{table}[tbp]
\centering\small
\setlength{\tabcolsep}{5pt}
\caption{Experiment 2: $r=1/2$ and $W=J_{\theta_0}^{(N)}$. The bound $\underline c_1$ is a risk lower bound valid for all POVMs, and $\overline c_{\mathrm{NH}}$ is an upper bound on the NH value.}\label{tab:experiment2-risk}
\begin{tabular}{@{}crrrr@{}}
\toprule
$N$ & $s$ & $c_{\mathrm{num}}$ & $\underline c_1$ (down) & $\overline c_{\mathrm{NH}}$ (up)\\
\midrule
2 & 8 & 6.000000000000 & 5.999998513757 & 6.000000553380\\
3 & 16 & 5.400000000000 & 5.399999310065 & 5.400000209279\\
4 & 32 & 5.014444831516 & 5.014444555033 & 5.014295965931\\
5 & 64 & 4.807929465702 & 4.807928216008 & 4.807806298074\\
6 & 128 & 4.658303399802 & 4.658301225392 & 4.657913754315\\
\bottomrule
\end{tabular}
\end{table}

\subsection{Experiment 3: General weights for two-copy qubit estimation}\label{sec:experiment3}
For $N=2$ in Eq.~\eqref{eq:experiment-qubit-model}, we consider two examples with weights expressed in standard Bloch coordinates:
\begin{equation}\label{eq:experiment3-models}
 \begin{aligned}
 \text{Diagonal}:\quad&r=\frac7{10},&
 W&=\frac1{1561}\diag(153,1530,3000),\\
 \text{Nondiagonal}:\quad&r=\frac9{100},&
 W&=\frac3{758}\begin{pmatrix}600&0&193\\0&92&0\\193&0&66\end{pmatrix}.
 \end{aligned}
\end{equation}
Both weights are positive definite and satisfy $\Tr W=3$.
The reported cost is $\Tr(WF^{-1})$ per batch of two copies, without the additional factor of two used in the reporting convention of Experiments 1 and 2.
For the diagonal example, we performed the computation in SLD-normalized coordinates $\theta=\theta_0+T\eta$ using
\[
 T=\diag(1,1,\sqrt{51}/10),\qquad
 W_{\mathrm{SLD}}=\frac17\diag(1,10,10).
\]
Because the weight in Eq.~\eqref{eq:experiment3-models} satisfies $W=(153/223)T^{-1}W_{\mathrm{SLD}}T^{-1}$, we converted the cost and both certified values in these coordinates by multiplying them by the common factor $153/223$.

We fixed the coefficient $B$ using a candidate measurement in the whitened coordinates and solved the $m=1$ SOS problem without Schur reduction.
Since $n=4,d=3$, the Gram matrix has order $4\binom{5}{2}=40$.
For the diagonal example, the reported measurement has $s=24$, whereas a separately obtained measurement with $s=16$ was used to determine the dual coefficient.
For the nondiagonal example, the same $s=32$ measurement was used both as the reported candidate and as the dual input.
Neither lower bound imposes any outcome-count restriction on the POVMs to which it applies.

Table~\ref{tab:experiment3-risk} establishes $c_{\mathrm{NH}}(W)\le \overline c_{\mathrm{NH}}<\underline c_1\le c_{\theta_0}(W)$ for both examples.
Rounding the guaranteed gaps conservatively gives
\[
 c_{\theta_0}(W)-c_{\mathrm{NH}}(W)\ge
 \begin{cases}
 2.0317\times10^{-7},&\text{diagonal example},\\
 6.2418\times10^{-6},&\text{nondiagonal example}
 \end{cases}.
\]
Thus, for suitable weights, the NH bound is unattainable even with two copies.
The separations from the numerical candidates, $c_{\mathrm{num}}-\underline c_1$, are approximately $1.63\times10^{-7}$ and $9.92\times10^{-8}$, respectively.
For both examples, the best optimization run terminated because the line search failed, and the SDP status was \texttt{optimal\_inaccurate}.
The strict separation above follows from the corrected rational certificates, not from the termination status.
\begin{table}[tbp]
\centering\small
\setlength{\tabcolsep}{5pt}
\caption{Experiment 3: Two copies, with $\Tr W=3$ in standard Bloch coordinates.}\label{tab:experiment3-risk}
\begin{tabular}{@{}lrrrr@{}}
\toprule
Weight & $s$ & $c_{\mathrm{num}}$ & $\underline c_1$ (down) & $\overline c_{\mathrm{NH}}$ (up)\\
\midrule
Diagonal & 24 & 1.792626688097 & 1.792626525116 & 1.792626321943\\
Nondiagonal & 32 & 2.116607707108 & 2.116607607878 & 2.116601365990\\
\bottomrule
\end{tabular}
\end{table}

\subsection{Experiment 4: Full eight-parameter qutrit model}\label{sec:experiment4}
Let $\mathcal H=\C^3$ and $\theta_0=0$, and define the local model by $\rho_\theta=\rho_{\theta_0}+\sum_{i=1}^8\theta^iD_i$.
In the standard basis $\{\ket0,\ket1,\ket2\}$, set $E_{ab}:=\ket a\bra b$ and,
for $(a_k,b_k)=(0,1),(0,2),(1,2)$ with $k=1,2,3$, define
\begin{equation}\label{eq:experiment4-basis}
 \begin{gathered}
 D_{2k-1}:=\frac{E_{a_kb_k}+E_{b_ka_k}}{\sqrt2},\qquad
 D_{2k}:=\frac{-iE_{a_kb_k}+iE_{b_ka_k}}{\sqrt2}\quad(k=1,2,3),\\
 D_7:=\frac{\diag(0,1,-1)}{\sqrt2},\qquad
 D_8:=\frac{\diag(2,-1,-1)}{\sqrt6}.
 \end{gathered}
\end{equation}
In this ordering, the basis satisfies $\tr(D_iD_j)=\delta_{ij}$.
We consider the two examples
\begin{equation}\label{eq:experiment4-models}
 \begin{aligned}
 \text{Example A}:\quad&\rho_{\theta_0}=\diag(14/15,1/30,1/30),&
 W&=\frac8{1281}\diag(20I_4,400I_3,1),\\
 \text{Example B}:\quad&\rho_{\theta_0}=\diag(4/5,1/10,1/10),&
 W&=\frac8{381}\diag(20I_4,100I_3,1).
 \end{aligned}
\end{equation}
Both satisfy $\Tr W=8$. The cost is $\Tr(WF^{-1})$ per qutrit.

Measurement optimization used $s=48$. On the SOS side, let $\widehat B=\widehat F^{-1}$ be the inverse Fisher matrix of the numerical measurement in coordinates that whiten the weight.
For the index sets $\mathcal J_1=\{1,2,3,4\}$, $\mathcal J_2=\{5,6,7\}$, and $\mathcal J_3=\{8\}$, we used the fixed coefficient
\begin{equation}\label{eq:experiment4-symmetric-B}
 B_{\mathrm{sym}}:=\diag\left(
 \frac{\Tr\widehat B_{\mathcal J_1,\mathcal J_1}}4I_4,\quad
 \frac{\Tr\widehat B_{\mathcal J_2,\mathcal J_2}}3I_3,\quad\widehat B_{88}\right).
\end{equation}
Here $\widehat B_{\mathcal J_a,\mathcal J_a}$ denotes the corresponding principal submatrix.
This operation selects the coefficient $B$ and does not modify the measurement itself.

The full Gram matrix at $m=1$ has order $3\binom{10}{2}=135$.
Using the sign symmetry $\mathbb Z_2^2$ of the problem with fixed $B_{\mathrm{sym}}$, we split the Gram matrix into four blocks of orders $35,35,35,30$.
Here $g=(g_1,g_2)\in\mathbb Z_2^2$ acts on the Hilbert space and on the polynomial variables $\xi=(\xi_1,\ldots,\xi_8)^{\mathsf T}\in\R^8$ as
\[
 \begin{aligned}
 D_g&:=\diag\bigl(1,(-1)^{g_1},(-1)^{g_2}\bigr),\\
 g\cdot\xi&:=\diag\bigl((-1)^{g_1}I_2,(-1)^{g_2}I_2,
                 (-1)^{g_1+g_2}I_2,I_2\bigr)\xi.
 \end{aligned}
\]
Define the characters by $\chi_c(g):=(-1)^{c_1g_1+c_2g_2}$ for $c=(c_1,c_2)\in\mathbb Z_2^2$, and set
$\eta_0=(0,0)$, $\eta_1=(1,0)$, and $\eta_2=(0,1)$.
The character of the Gram coordinate $\xi^\alpha\ket a$, with $\alpha\in\mathbb N_0^8$, $|\alpha|\le2$, and $a=0,1,2$, is
\[
 c(a,\alpha):=\eta_a+
 \bigl(\alpha_1+\alpha_2+\alpha_5+\alpha_6,\,
       \alpha_3+\alpha_4+\alpha_5+\alpha_6\bigr)\pmod2.
\]
Averaging over the four group elements eliminates Gram entries between different characters, so coordinates with the same $c(a,\alpha)$ form a block.
Together with $X\mapsto D_gXD_g^\dagger$, this group average preserves feasibility and the objective value of the same SDP with fixed $B_{\mathrm{sym}}$. Imposing the block structure therefore leaves its optimum unchanged.
This decomposition acts on the Gram space of monomials and Hilbert-space components; it differs from the Hilbert-space decomposition by a sufficient subalgebra in Sec.~\ref{sec:sos-sufficiency}.
The coefficient equalities couple the four blocks, so the problem does not split into four independent SDPs.
We also do not assume that replacing $\widehat B$ by $B_{\mathrm{sym}}$ preserves the optimum of the fixed-coefficient problem.

Table~\ref{tab:experiment4-risk} establishes that the NH bound is unattainable in both full eight-parameter examples.
Including corrections to the ideal state, the square roots in the basis, and the weight, we rigorously obtain
\begin{equation}\label{eq:experiment4-nh-gap}
 c_{\theta_0}(W)-c_{\mathrm{NH}}(W)\ge
 \begin{cases}
 0.306019099900,&\text{example A},\\
 0.642113385162,&\text{example B}
 \end{cases}.
\end{equation}
These bounds guarantee relative gaps with respect to the NH value of at least $14.620419\%$ for example A and $11.089265\%$ for example B.
The relative-gap lower bounds were obtained by evaluating $(\underline c_1-\overline c_{\mathrm{NH}})/\overline c_{\mathrm{NH}}$ downward from the rational values before rounding.
The lower bounds for all POVMs impose no restrictions on the outcome count, rank, or symmetry of the measurement.

The level-$m=1$ certified lower bounds differ from the best numerical measurement costs by only $1.09\times10^{-6}$ and $1.39\times10^{-5}$ for examples A and B, respectively.
The corresponding relative separations, $(c_{\mathrm{num}}-\underline c_1)/c_{\mathrm{num}}$, are approximately $4.54\times10^{-7}$ and $2.16\times10^{-6}$.
Thus, even in examples with certified gaps exceeding $14\%$ and $11\%$ from the NH bound, this low level provides numerical evidence of near-optimality.
This accuracy is obtained with fixed support coefficients, without joint optimization over $B$.

All 10 runs in each example terminated because the line search failed; the closeness of the costs across runs is not taken as evidence of gradient convergence.
Example A includes very small effects, with traces as low as approximately $2.08\times10^{-20}$. These effects were not removed, and all 48 rows of the resulting Stiefel matrix are counted as outcomes.
\begin{table}[tbp]
\centering\small
\setlength{\tabcolsep}{5pt}
\caption{Experiment 4: One qutrit, with $\Tr W=8$. The gaps in Eq.~\eqref{eq:experiment4-nh-gap} are independently rounded downward from the rational values before display.}\label{tab:experiment4-risk}
\begin{tabular}{@{}lrrrr@{}}
\toprule
Example & $s$ & $c_{\mathrm{num}}$ & $\underline c_1$ (down) & $\overline c_{\mathrm{NH}}$ (up)\\
\midrule
A & 48 & 2.399113962354 & 2.399112871985 & 2.093093772085\\
B & 48 & 6.432532266367 & 6.432518351067 & 5.790404965906\\
\bottomrule
\end{tabular}
\end{table}

\subsection{Computational efficiency of the two-stage optimization}\label{sec:barc-evidence}
We compared the performance of the two-stage optimization for $N=4$ identical, independent qubits, with $n=16$, $s=24$, and $r=0.001$.
With $W=J_{\theta_0}^{(4)}$ and the SLD-normalized coordinates of Sec.~\ref{sec:experiment1}, we report $c_{\mathrm{num}}=4\Tr\widetilde F^{-1}$; the optimization objective is $\Tr\widetilde F^{-1}$.
The Q-StARC timings in this subsection cover only measurement optimization and exclude the SOS lower-bound computation.

First, we evaluated the effect of appending RCG to Adam using five initializations.
RCG refinement reduced $c_{\mathrm{num}}$ for every initialization.
Adam took approximately $1.147$--$1.151$ s, and the additional RCG stage took approximately $0.0397$--$0.1084$ s, accounting for approximately $3.34$--$8.62\%$ of the total time.
For example, for seed 2, approximately $0.0397$ s of RCG reduced the cost as follows:
\[
 c_{\mathrm{num}}:\quad 4.965531939794\ \longrightarrow\ 4.965517412604.
\]
This supports the use of a short refinement stage after the initial search.
Of the five runs, four stopped at the iteration limit and one stopped because the line search failed; none terminated by satisfying the gradient stopping criterion.

Next, to compare the first-stage updates, we compared projected ambient Adam with Riemannian Adam, which transports the first moment by tangent-space projection and uses the squared norm of the full gradient for the second moment.
The comparator is an independent JAX implementation that extends the RiemannianAdam and EuclideanStiefel update formulas of Geoopt~0.5.0\cite{Geoopt050} to the complex Stiefel manifold. It shares the objective, metric, QR retraction, and subsequent RCG stage with the Q-StARC measurement optimizer.
For each method, we compared 10 candidate learning rates using five tuning initializations. After fixing the settings with the same selection rule, which prioritizes the best attained value, we evaluated them on 20 held-out initializations.
Each run consisted of $30000$ Adam-type updates followed by at most $2000$ RCG iterations, and each initialization was timed three times.

\begin{table}[tbp]
\centering\small
\setlength{\tabcolsep}{4pt}
\caption{Performance comparison with the same iteration budget. The reported time is the median over 20 initializations of the median of three timings per initialization.}\label{tab:barc-performance}
\begin{tabular}{@{}lrrrrr@{}}
\toprule
Method & Learning rate & Time (s) & Best $c_{\mathrm{num}}$ & Median $c_{\mathrm{num}}$ & Target hits\\
\midrule
Q-StARC (default) & $0.005$ & $1.3006$ & $4.971830910$ & $4.993696181$ & $0/20$\\
Q-StARC (tuned) & $0.1$ & $1.3013$ & $4.965517413$ & $4.971831137$ & $6/20$\\
Riemannian Adam--RCG & $0.07071$ & $1.2917$ & $4.993682875$ & $5.038310165$ & $0/20$\\
\bottomrule
\end{tabular}
\end{table}

We counted a target hit when the cost was at most the numerical reference value $4.965517412604129$ plus $10^{-8}$.
This reference value is not a rigorous lower bound, and the number of target hits is distinct from the number of runs satisfying the gradient stopping criterion.
The tuned Q-StARC measurement optimizer attained a lower $c_{\mathrm{num}}$ than the Riemannian Adam comparator for all 20 paired initializations.
All methods took approximately $1.3$ s, so tuned Q-StARC produced lower-cost measurement candidates in a comparable time.

The computations used double precision on a CPU; the environment for Table~\ref{tab:barc-performance} was macOS arm64 with JAX~0.11.2.
The timings exclude JIT compilation, generation of the model and initial measurement, and independent verification, but include optimizer-state initialization, synchronization for observation, and the transition between stages.
Learning-rate tuning took approximately $133$ s in total, which is excluded from the optimization times in the table.
In this setting, RCG refinement reduced the objective, and the two-stage method with a tuned learning rate attained a lower objective than the comparator in a comparable time.
This experiment does not separately evaluate the contributions of rank-one effects, direct Stiefel parametrization, and QR retraction to the speedup.

\subsection{Timings and scope of comparison in Experiments 2--4}\label{sec:experiment-times}
Table~\ref{tab:experiment-times} reports the computation times.
The measurement-optimization time is the total over the 10 initializations used, rather than the time for the single best run.
The optimization environment for Experiments 1 and 2 was macOS arm64, Python~3.12.14, NumPy~2.5.4, and JAX~0.11.2.
In addition to the dual SDP solve time, we report a computation time that includes problem construction and certification procedures.
\begin{table}[tbp]
\centering\small
\setlength{\tabcolsep}{5pt}
\caption{Computation times in Experiments 2--4, in seconds. Measurement optimization is totaled over 10 initializations. The scope of the SOS computation time is specified in the text.}\label{tab:experiment-times}
\begin{tabular}{@{}llrrrr@{}}
\toprule
Exp. & Case & Optimization & SOS solve & SOS computation & Separate warmup\\
\midrule
2 & $N=2$ & 1.931 & 1.080 & 1.109 & 0.955\\
  & $N=3$ & 4.249 & 5.734 & 5.782 & 0.657\\
  & $N=4$ & 16.023 & 20.533 & 20.619 & 0.711\\
  & $N=5$ & 94.156 & 57.657 & 57.791 & 0.793\\
  & $N=6$ & 375.654 & 124.292 & 124.479 & 0.717\\
\midrule
3 & Diagonal & 3.433 & 5.216 & 5.259 & ---\\
  & Nondiagonal & 3.833 & 4.485 & 4.519 & ---\\
\midrule
4 & A & 5.268 & 36.095 & 36.263 & ---\\
  & B & 5.258 & 34.586 & 34.777 & ---\\
\bottomrule
\end{tabular}
\par\smallskip\footnotesize
--- indicates that the warmup time could not be identified separately.
\end{table}

The optimization time in Experiment 2 excludes the initial JIT warmup.
Its SOS timing covers construction, conversion, SDP solution, and dyadic-rational certification for all Schur blocks, but excludes the additional correction to ideal inputs and independent reverification.
For Experiment 3, five stages were timed: construction, conversion, SDP solution, rational certification, and correction to ideal inputs. The reported total is the sum of these stage times, not the elapsed time from the start to the end of the entire procedure.
The separate 10-initialization optimization with $s=16$ used as the dual input for the diagonal example took $2.811$ s; it is not included in the $s=24$ optimization time in the table.
For Experiment 4, the timing covers construction, solution, and certification corrections. It excludes the aborted unsplit SDP computation, the full candidate search, and independent reverification.
The SDPs in Experiment 4 were solved with CLARABEL, using one thread and a cold start.
For Experiments 3 and 4, the JIT time incurred within optimization calls cannot be separated from the other work.
Concurrent workload was not controlled across experiments, and the timed operations differ. The ratios of times in the table therefore do not measure relative computational efficiency between experiments.
\section{Conclusion}\label{sec:conclusion}
We have developed a framework that combines efficient numerical measurement design with certified risk bounds for multiparameter quantum estimation.
Rank-one Stiefel optimization with projected ambient Adam, QR retraction, and RCG refinement produces measurement candidates with small weighted variance.
Independent lower bounds are obtained from a matrix-SOS hierarchy whose joint-coefficient formulation starts at the NH bound and converges to the optimal locally unbiased risk, including for singular states under the assumptions of Sec.~\ref{sec:model}.

The numerical results certify that the NH bound is unattainable for the SLD-weighted qubit examples with $N=4,5,6$, the two general-weight two-copy qubit examples, and the two qutrit examples.
For the qutrit examples, the certified gaps above the NH value exceed $14\%$ and $11\%$, while the level-$m=1$ bounds with fixed support coefficients agree with the best numerical measurement costs to relative differences of approximately $4.5\times10^{-7}$ and $2.2\times10^{-6}$.
These results show that low levels of the hierarchy can provide numerically near-tight bounds even when the NH relaxation is substantially below the attainable risk.
The exact risk of a feasible candidate measurement is an upper bound, but the reported floating-point values do not include rigorous bounds on errors in POVM construction and risk evaluation.
Thus this evidence of near-optimality is distinct from the certified NH nonattainability.
The optimizer comparison evaluates the complete two-stage method; it does not isolate the contribution of each component.

Sufficient-algebra reduction is already used in the numerical SOS calculations and can also be incorporated into measurement search through the blockwise Stiefel formulation presented here.
Further numerical work can assess the gains in bound tightness from joint coefficient optimization and higher SOS levels relative to their computational cost.
Together, these developments would strengthen the connection between practical measurement design and quantitative guarantees on attainable precision.
\section*{Acknowledgments}
This work was supported by JSPS KAKENHI Grant Numbers JP23H01090 and
JP22K03466 and JST ERATO Grant Number JPMJER2402, Japan.
\section*{Data availability}
A public implementation of Q-StARC is available on GitHub as version 1.0.1~\cite{QStARC}.
The release includes measurement optimization, SOS lower-bound computation for fixed support coefficients $B$, certificate verification, and runnable examples.
The scope of the release and its certification guarantees are documented in the repository README.
The release does not include the complete experimental data or the experiment-specific input-error corrections used for the certified bounds reported here.
The numerical data underlying the figures and tables, together with the additional code and certificate-verification files used in the reported experiments, are available from the authors upon reasonable request.
\appendix
\renewcommand{\thetheorem}{\Alph{section}.\arabic{theorem}}
\renewcommand{\theequation}{\Alph{section}.\arabic{equation}}
\section{NH and Holevo bounds and the initial Gram relaxation}\label{app:nh}
This appendix applies SDP duality to the NH primal problem and identifies its dual with the initial Gram relaxation. Relaxing the symmetry constraints in the primal gives the Holevo bound, which corresponds to a restriction of the Gram matrix on the dual side.
We work on the original Hilbert space $\mathcal H$ and write $n=\dim\mathcal H$, $\rho=\rho_{\theta_0}$, $D_j=\partial_j\rho_{\theta_0}$, and $S=W^{1/2}$.

\subsection{A common SDP duality formula}\label{app:sdp-duality}
Let $\mathcal E,\mathcal Y$ be finite-dimensional real inner-product spaces and let
$\mathsf A:\mathcal E\to\mathcal Y$ and $\mathsf F:\mathcal E\to\mathcal B_h(\C^N)$ be linear maps. Consider
\[
 p_*:=\inf_{x\in\mathcal E}
 \{\langle c,x\rangle:\mathsf Ax=b,\quad K(x):=F_0+\mathsf Fx\ge0\}
\]
where $c\in\mathcal E$, $b\in\mathcal Y$, and $F_0\in\mathcal B_h(\C^N)$ are fixed.
On the space of self-adjoint matrices, use the inner product $\langle U,V\rangle=\tr(UV)$ and define the adjoint maps by
\[
 \langle y,\mathsf Ax\rangle=\langle\mathsf A^{\dagger}y,x\rangle,
 \qquad
 \langle G,\mathsf Fx\rangle=\langle\mathsf F^{\dagger}G,x\rangle
\]
Associate $y\in\mathcal Y$ with the equality constraint and $G\ge0$ with the semidefinite constraint. The Lagrangian is
\[
 \begin{aligned}
 \mathfrak L(x;y,G)
 &=\langle c,x\rangle-\langle y,\mathsf Ax-b\rangle
   -\langle G,F_0+\mathsf Fx\rangle\\
 &=\langle b,y\rangle-\langle F_0,G\rangle
   +\langle c-\mathsf A^{\dagger}y-\mathsf F^{\dagger}G,x\rangle.
 \end{aligned}
\]
For fixed $y,G$, remove the original constraints and take the infimum over all $x\in\mathcal E$.
If $H:=c-\mathsf A^{\dagger}y-\mathsf F^{\dagger}G\ne0$, then
$\langle H,-tH\rangle=-t\|H\|^2\to-\infty$. Hence
\[
 \inf_x\mathfrak L(x;y,G)=
 \begin{cases}
 \langle b,y\rangle-\langle F_0,G\rangle,
   &\mathsf A^{\dagger}y+\mathsf F^{\dagger}G=c,\\
 -\infty,&\text{otherwise}.
 \end{cases}
\]
The dual problem is therefore
\begin{equation}\label{eq:general-sdp-dual}
 d_*:=\sup_{y,\,G\ge0}
 \{\langle b,y\rangle-\langle F_0,G\rangle:
   \mathsf A^{\dagger}y+\mathsf F^{\dagger}G=c\}.
\end{equation}
Thus the dual constraints follow by comparing the coefficients of the independent variable $x$.
At primal and dual feasible points,
\[
 \langle c,x\rangle-
 \{\langle b,y\rangle-\langle F_0,G\rangle\}
 =\tr\bigl(GK(x)\bigr)\ge0
\]
gives $d_*\le p_*$. If, in addition, $p_*$ is finite and there exists $x_0$ with $\mathsf Ax_0=b$ and $K(x_0)>0$, Slater's condition gives $p_*=d_*$ and attainment of the dual optimum~\cite{BoydVandenberghe}.

\subsection{Application to the NH primal problem}
For a locally unbiased measurement and estimator, define the weighted error $a(x):=S(\hat\theta(x)-\theta_0)$ and set
\[
 R_i:=\sum_xa_i(x)M_x,\qquad
 \tr(\rho R_i)=0,\qquad
 \tr(D_jR_i)=\sum_kS_{ik}\delta^k_j=S_{ij}
\]
In the optimization below, regard $R_i\in\mathcal B_h(\mathcal H)$ as independent variables and write
\[
 \mathbf R:=\begin{pmatrix}R_1\\\vdots\\R_d\end{pmatrix}:
 \mathcal H\longrightarrow\C^d\otimes\mathcal H,\qquad
 \mathcal L=(\mathcal L_{ij})_{i,j=1}^d\in\mathcal B_h(\C^d\otimes\mathcal H)
\]
The NH bound is~\cite{NH,HO}
\begin{equation}\label{eq:nh-primal}
\begin{aligned}
 c_{\mathrm{NH}}(W)=\inf_{R,\mathcal L}\quad
 &f(\mathcal L):=\sum_i\tr(\rho\mathcal L_{ii})\\
 \text{subject to}\quad &\mathbf M(R,\mathcal L):=
 \begin{pmatrix}I_n&\mathbf R^{\dagger}\\\mathbf R&\mathcal L\end{pmatrix}\ge0,\\
 &\tr(D_jR_i)=S_{ij},\qquad
 \mathcal L_{ij}=\mathcal L_{ji}=\mathcal L_{ij}^{\dagger}
 \quad(1\le i,j\le d)
\end{aligned}
\end{equation}
The omission of the zero-mean constraints is justified below by centering.
The second moments $\mathcal L_{ij}:=\sum_xa_i(x)a_j(x)M_x$ satisfy
\[
 \mathbf M=\sum_x\begin{pmatrix}1\\a(x)\end{pmatrix}
       \begin{pmatrix}1&a(x)^{\mathsf T}\end{pmatrix}\otimes M_x\ge0,
 \qquad f(\mathcal L)=\Tr\bigl(WV_{\theta_0}[M,\hat\theta]\bigr)
\]
so Eq.~\eqref{eq:nh-primal} gives a lower bound on the estimation risk.

\paragraph{Independent variables and the duality formula}
The blocks $\mathcal L_{ij}$ and $\mathcal L_{ji}$ are the same self-adjoint variable. Accordingly, take
\[
 \begin{gathered}
 x:=\bigl((R_i)_i,(\mathcal L_{ii})_i,(\mathcal L_{ij})_{i<j}\bigr),\\
 \langle x,\widetilde x\rangle
 :=\sum_i\tr(R_i\widetilde R_i)
   +\sum_i\tr(\mathcal L_{ii}\widetilde{\mathcal L}_{ii})
   +\sum_{i<j}\tr(\mathcal L_{ij}\widetilde{\mathcal L}_{ij})
 \end{gathered}
\]
counting each off-diagonal variable only once. Substitute
\[
 \begin{gathered}
 (\mathsf Ax)_{ij}:=\tr(D_jR_i),\qquad b=S,\qquad
 \langle c,x\rangle=f(\mathcal L),\\
 F_0:=\begin{pmatrix}I_n&0\\0&0\end{pmatrix},\qquad
 \mathsf Fx:=\begin{pmatrix}0&\mathbf R^{\dagger}\\\mathbf R&\mathcal L\end{pmatrix},
 \qquad K(x)=\mathbf M(R,\mathcal L)
 \end{gathered}
\]
into the duality formula above.
Write the equality multiplier as $y=2B$, with $B\in\R^{d\times d}$, and set $A_i:=\sum_jB_{ij}D_j$.
With the real-matrix inner product $\langle U,V\rangle=\Tr(U^{\mathsf T}V)$, we obtain
\[
 \begin{aligned}
 \langle2B,\mathsf Ax\rangle
 &=\sum_i\tr(2A_iR_i),\\
 \langle G,\mathsf Fx\rangle
 &=\sum_i\tr\bigl[(G_{0i}+G_{i0})R_i\bigr]
   +\sum_i\tr(G_{ii}\mathcal L_{ii})
   +\sum_{i<j}\tr\bigl[(G_{ij}+G_{ji})\mathcal L_{ij}\bigr].
 \end{aligned}
\]
Here $G=(G_{ij})_{i,j=0}^d\ge0$ and $G_{ij}\in\mathcal B(\mathcal H)$.
The coefficients of the independent variables give the components of the adjoints:
\[
 \begin{array}{c|ccc}
 \text{Variable}&c&\mathsf A^{\dagger}(2B)&\mathsf F^{\dagger}G\\\hline
 R_i&0&2A_i&G_{0i}+G_{i0}\\
 \mathcal L_{ii}&\rho&0&G_{ii}\\
 \mathcal L_{ij}\ (i<j)&0&0&G_{ij}+G_{ji}
 \end{array}
\]
Equating the sum of the last two columns to the $c$ column and using
$\langle S,2B\rangle-\langle F_0,G\rangle=2\Tr(BS)-\tr G_{00}$ yields the following dual representation.

\begin{theorem}[NH dual and the initial SOS relaxation]\label{thm:nh-dual-sos}
Consider the quantum model of Sec.~\ref{sec:model} and a positive-definite weight $W$.
For the variables $B\in\R^{d\times d}$ and
$G=(G_{ij})_{i,j=0}^d\in\mathcal B_h(\C^{d+1}\otimes\mathcal H)$,
set $S=W^{1/2}$ and $A_i=\sum_jB_{ij}D_j$. Then
\[
 \begin{aligned}
 c_{\mathrm{NH}}(W)=c_0(W)=\max_{B,G}\quad&2\Tr(BS)-\tr G_{00}\\
 \text{subject to}\quad&G\ge0,\qquad G_{ii}=\rho\quad(1\le i\le d),\\
 &G_{0i}+G_{i0}=-2A_i\quad(1\le i\le d),\\
 &G_{ij}+G_{ji}=0\quad(1\le i<j\le d)
 \end{aligned}
\]
Here $c_0(W)$ is the zeroth level of the SOS hierarchy with joint coefficient optimization in Eq.~\eqref{eq:joint-sdp}.
The maximum is attained, and the statements remain valid when $\rho$ is singular.
\end{theorem}
The dual constraints follow from the coefficient comparison above. Equality of the primal and dual values and attainment of the maximum are established in Appendix~\ref{app:nh-strong-duality}; the equivalence with the initial SOS relaxation is proved in Appendix~\ref{app:nh-sos}.
Although $G$ is self-adjoint as a whole, its off-diagonal blocks are not individually constrained to be self-adjoint.
For example, the last condition is $G_{ij}+G_{ij}^{\dagger}=0$, leaving the anti-self-adjoint part unrestricted.

\subsection{Equality of the primal and dual values}\label{app:nh-strong-duality}
Let $(R,\mathcal L)$ be the NH feasible point formed from the moments of the finite-risk locally unbiased estimator obtained from Proposition~\ref{prop:sld-randomization} and Eq.~\eqref{eq:efficient-estimator}.
For any $\varepsilon>0$, setting $\mathcal L^{(\varepsilon)}:=\mathcal L+\varepsilon I_{dn}$ preserves the equality constraints.
The Schur complement gives
\[
 \mathbf M(R,\mathcal L)\ge0
 \quad\Longrightarrow\quad
 \mathcal L-\mathbf R\mathbf R^{\dagger}\ge0
\]
and hence
\[
 \mathcal L^{(\varepsilon)}-\mathbf R\mathbf R^{\dagger}
 \ge\varepsilon I_{dn}>0,
 \qquad \mathbf M(R,\mathcal L^{(\varepsilon)})>0.
\]
Moreover,
\[
 0\le c_{\mathrm{NH}}(W)\le f(\mathcal L^{(\varepsilon)})
 =f(\mathcal L)+d\varepsilon<\infty.
\]
The Slater condition in Appendix~\ref{app:sdp-duality} therefore applies: the primal and dual values coincide, and the dual optimum is attained.
This proves the maximum representation in Theorem~\ref{thm:nh-dual-sos}.

\subsection{Equivalence with the initial SOS and Gram problems}\label{app:nh-sos}
With $X=G_{00}$, the NH dual constraints are exactly those in Eq.~\eqref{eq:initial-gram}.
Compression to the common effective space of Sec.~\ref{sec:effective-space} preserves $d_0(B)$ for every fixed $B$.
Applying Lemma~\ref{lem:graded-degree}(i) with $m=0$ on that space gives
\[
 Q_{B,X}\text{ is a matrix SOS}
 \quad\Longleftrightarrow\quad
 \exists\,\widetilde G\ge0:\quad
 Q_{B,X}=V_0\widetilde G V_0^{\dagger}.
\]
This equivalence holds for the same $X$, so it also preserves the objective value.
The strong duality proved above and the definition in Eq.~\eqref{eq:joint-sdp} therefore imply
\[
 c_{\mathrm{NH}}(W)
 =\sup_{B\in\R^{d\times d}}\{2\Tr(BS)-d_0(B)\}=c_0(W)
\]
This completes the proof of Theorem~\ref{thm:nh-dual-sos}. In particular, for every fixed $B$,
\[
 c_0(W;B)=2\Tr(BS)-d_0(B)\le c_{\mathrm{NH}}(W).
\]

\paragraph{Centering and the zero-mean constraints}
We now justify the omission of the zero-mean constraints in Eq.~\eqref{eq:nh-primal}.
For $\mu_i:=\tr(\rho R_i)$ and $\mu=(\mu_i)_{i=1}^d\in\R^d$, set
\[
 \begin{aligned}
 R_i'&:=R_i-\mu_iI_n,\\
 \mathcal L_{ij}'&:=\mathcal L_{ij}-\mu_iR_j-\mu_jR_i+\mu_i\mu_jI_n,\\
 C_\mu&:=\begin{pmatrix}1&0\\-\mu&I_d\end{pmatrix}\otimes I_n,\qquad
 \mathbf M(R',\mathcal L')=C_\mu\mathbf M(R,\mathcal L)C_\mu^{\dagger}\ge0.
 \end{aligned}
\]
Since $\tr\rho=1$ and $\tr D_j=0$,
\[
 \begin{gathered}
 \tr(\rho R_i')=0,\qquad \tr(D_jR_i')=S_{ij},\qquad
 \mathcal L_{ij}'=\mathcal L_{ji}'=(\mathcal L_{ij}')^{\dagger},\\
 \sum_i\tr(\rho\mathcal L_{ii}')
 =\sum_i\tr(\rho\mathcal L_{ii})-\|\mu\|^2.
 \end{gathered}
\]
Thus centering preserves feasibility and does not increase the objective.

\subsection{The Gram restriction yielding the Holevo bound}\label{app:holevo}
For the same weighted variables $R$, define $Z(R)_{ij}:=\tr(\rho R_iR_j)$.
Here $\operatorname{Re}$ and $\operatorname{Im}$ denote entrywise real and imaginary parts.
The Holevo bound is~\cite{AFD}
\begin{equation}\label{eq:holevo}
 c_{\mathrm H}(W)=\inf_{R,V}\{\Tr V:V\in\mathcal{B}_{hr}^d,\quad
 V\ge Z(R),\quad\tr(D_jR_i)=S_{ij}\}
\end{equation}

\begin{theorem}[Dual of the Holevo bound]\label{thm:holevo-dual}
Under the assumptions of Theorem~\ref{thm:nh-dual-sos}, use the same variables $B,G$ and
an additional variable $K\in\mathcal B_h(\C^d)$. Then
\begin{equation}\label{eq:holevo-gram}
 \begin{aligned}
 c_{\mathrm H}(W)=\max_{B,G,K}\quad&2\Tr(BS)-\tr G_{00}\\
 \text{subject to}\quad&G\ge0,\quad G_{0i}+G_{i0}=-2A_i\quad(1\le i\le d),\\
 &G_{ij}=K_{ij}\rho\quad(1\le i,j\le d),\\
 &K\ge0,\qquad\operatorname{Re}K=I_d
 \end{aligned}
\end{equation}
The maximum is attained. This is the NH dual with its operator blocks indexed by $i,j\ge1$
restricted to $G_{ij}=K_{ij}\rho$, and it remains valid for singular $\rho$.
\end{theorem}

\begin{proof}
\par\medskip\noindent\textit{Relaxation of the NH primal.}\enspace
Equation~\eqref{eq:holevo} is equivalent to relaxing the NH primal constraints $\mathcal L_{ij}=\mathcal L_{ji}$ to
$\operatorname{Im}\tr(\rho\mathcal L_{ij})=0$ for $i<j$.
Indeed, the positive map on operator-block matrices $Y=(Y_{ij})$ defined by
\[
 \Phi_\rho(Y):=\bigl(\tr(\rho Y_{ij})\bigr)_{i,j=1}^d
\]
satisfies
\[
 \Phi_\rho(\mathbf R\mathbf R^{\dagger})=Z(R),\qquad
 \Phi_\rho(C\otimes I_n)=C\quad\bigl(C\in\mathcal B_h(\C^d)\bigr)
\]
The Schur complement therefore gives, for $V\in\mathcal B_h(\C^d)$,
\[
 \exists\,\mathcal L:\quad
 \mathbf M(R,\mathcal L)\ge0,\quad\Phi_\rho(\mathcal L)=V
 \quad\Longleftrightarrow\quad V\ge Z(R).
\]
For the reverse implication, choose
\[
 \mathcal L=\mathbf R\mathbf R^{\dagger}+(V-Z(R))\otimes I_n
\]
The real-valuedness constraints are equivalent to $V\in\mathcal B_{hr}^d$. Since
$\Tr V=f(\mathcal L)$, the two optimal values coincide.
The centering above preserves real-valuedness because $\tr(\rho\mathcal L'_{ij})=\tr(\rho\mathcal L_{ij})-\mu_i\mu_j$; hence the zero-mean constraints can again be imposed without increasing the objective.

\par\medskip\noindent\textit{Derivation of the Holevo dual.}\enspace
The coefficient constraints for $R_i$ and the dual objective are the same as in the NH case; only the constraints associated with $\mathcal L$ change.
Regard $\mathcal L\in\mathcal B_h(\C^d\otimes\mathcal H)$ as a free variable. With a real antisymmetric matrix $\Omega\in\R^{d\times d}$, add the term
\[
 2\sum_{i<j}\Omega_{ij}\operatorname{Im}\tr(\rho\mathcal L_{ij})
 =\tr\bigl[(i\Omega\otimes\rho)\mathcal L\bigr],
 \qquad\Omega^{\mathsf T}=-\Omega
\]
to the Lagrangian for the real-valuedness constraints. Writing $G_{++}:=(G_{ij})_{i,j=1}^d$, the part depending on $\mathcal L$ is
\[
 \tr\bigl[\bigl((I_d+i\Omega)\otimes\rho-G_{++}\bigr)\mathcal L\bigr]
\]
Its infimum is finite if and only if
\[
 G_{++}=K\otimes\rho,\qquad K:=I_d+i\Omega.
\]
Furthermore, $\tr\rho=1$ and $G_{++}\ge0$ imply
\[
 K=\tr_{\mathcal H}G_{++}\ge0,\qquad\operatorname{Re}K=I_d.
\]
Together with the constraints shared with the NH case, these are the dual constraints in Eq.~\eqref{eq:holevo-gram}.
From $K=K^{\dagger}$ and $\operatorname{Re}K=I_d$,
\[
 G_{ii}=K_{ii}\rho=\rho,\qquad
 G_{ij}+G_{ji}=(K_{ij}+K_{ji})\rho=0\quad(i<j).
\]
All other constraints and the objective are unchanged. Thus the Holevo dual is the restriction of the NH dual to $G_{ij}=K_{ij}\rho$ for $i,j\ge1$.

The strictly feasible NH point constructed in Appendix~\ref{app:nh-strong-duality} is also strictly feasible for this relaxed primal. Moreover, the relaxation gives
\[
 0\le c_{\mathrm H}(W)\le c_{\mathrm{NH}}(W)<\infty
\]
so the Slater condition in Appendix~\ref{app:sdp-duality} proves equality of the primal and dual values and attainment of the dual optimum.
\end{proof}

Theorems~\ref{thm:nh-dual-sos} and~\ref{thm:holevo-dual}, together with monotonicity of the hierarchy, give
\[
 c_{\mathrm H}(W)\le c_{\mathrm{NH}}(W)=c_0(W)\le c_1(W)\le\cdots\le c_{\theta_0}(W).
\]
This representation expresses the abstract conic dual of Hayashi and Ouyang~\cite{HO} as explicit semidefinite constraints on operator blocks.
It identifies the Holevo dual as a block restriction of the NH dual and makes their relation to the initial SOS relaxation explicit in a common Gram variable.
\section{A bound on the number of outcomes in each block}\label{app:block-support}
\begin{theorem}[Blockwise outcome bound]\label{thm:block-support}
Under the assumptions of Sec.~\ref{sec:barc-sufficiency}, let $s_b^{\max}:=n_b^2+d(d+1)/2-1$.
Any POVM of the form \eqref{eq:barc-product-stiefel} can be replaced by one with $n_b\le s_b\le s_b^{\max}$ in each block, without decreasing the blockwise Fisher information, while preserving completeness and the rank-one property of every nonzero effect on the reduced block.
Moreover,
\[
 c_{\theta_0}(W)=\inf\left\{f_W(\boldsymbol Z):
 \boldsymbol Z\in\prod_{b=1}^{q}\mathrm{St}_{\C}(s_b^{\max},n_b),\quad
 F(\boldsymbol Z)>0\right\}
\]
holds with the number of outcomes in each block fixed at its upper bound. Thus, no search over outcome allocations is required.
\end{theorem}
\begin{proof}
Fix a block $b$. In the notation of \eqref{eq:fisher},
\[
 F_b:=F_b(Z_b)=\sum_xg_{\theta_0}[M_{b,x}],\qquad
 g_{\theta_0}[cM_{b,x}]=c\,g_{\theta_0}[M_{b,x}]
 \quad(c\ge0).
\]
The outcome-elimination argument of Yamagata~\cite{Yamagata}, applied to the real dimension $n_b^2$ of the Hermitian operator space on the reduced block, gives $s_b\le s_b^{\max}$ without decreasing $F_b$, while preserving completeness and rank-one effects on that block.
This argument does not use normalization of the state and therefore also applies to the unnormalized $\rho_b$.

Applying this reduction to each block does not decrease the total Fisher information and hence does not increase the risk.
Finally, splitting a nonzero effect $M_{b,x}$ into $a=s_b^{\max}-s_b+1$ copies of $M_{b,x}/a$ gives exactly $s_b^{\max}$ outcomes without changing the Fisher information, by positive homogeneity.
The reduction using sufficiency in Sec.~\ref{sec:barc-sufficiency} and the preceding operations do not increase the risk. Conversely, every measurement on the fixed product Stiefel manifold is feasible for the original problem. The two infima therefore coincide.
\end{proof}
\section{Finite Vidal measurement in Experiment 1}\label{app:finite-vidal}
We specify the directions and weights of the finite Vidal measurement compared in Sec.~\ref{sec:experiment1}.
For $q=2,\ldots,6$, we use the constructions of Latorre, Pascual, and Tarrach\cite{LPT} in their original orientations.
We parametrize the directions as
\[
 n(z,\phi):=\bigl(\sqrt{1-z^2}\cos\phi,\sqrt{1-z^2}\sin\phi,z\bigr),
 \qquad -1\le z\le1,\quad 0\le\phi<2\pi,
\]
and choose $\phi=0$ at $z=\pm1$.
With the Pauli-matrix convention of Sec.~\ref{sec:experiment1}, the corresponding states are
\[
 \ket{n(z,\phi)}
 :=\sqrt{\frac{1+z}{2}}\ket0
   +e^{i\phi}\sqrt{\frac{1-z}{2}}\ket1,\qquad
 \ket{q/2,n(z,\phi)}:=\ket{n(z,\phi)}^{\otimes q}
 \quad(q\ge1).
\]
For $q=0$, we set $\operatorname{Sym}^0(\C^2)=\C$ and take the state to be $1$, with no direction assigned.
In Table~\ref{tab:finite-vidal-directions}, $c_{q,x}^2$ is the weight of the effect
$c_{q,x}^2\ket{q/2,n_{q,x}}\bra{q/2,n_{q,x}}$ associated with each direction; it does not include the multiplicity $\mu_{N,q}$.
The constants used in the table are
\[
 z_\pm:=\frac{\sqrt{13\pm2\sqrt{30}}}{7},\qquad
 w_\pm:=\frac{7(410\pm\sqrt{30})}{7200}.
\]

\begin{table}[tbp]
\centering\small
\setlength{\tabcolsep}{10pt}
\renewcommand{\arraystretch}{1.00}
\caption{All directions and weights fixed in Experiment 1. Each value of $k$ and each choice of the $\pm$ sign in a row defines a separate outcome $x$.}\label{tab:finite-vidal-directions}
\begin{tabular}{@{}ccccc@{}}
\toprule
$q$ & $z$ & $\phi/\pi$ & Range of $k$ & $c_{q,x}^2$\\
\midrule
0 & No direction & --- & --- & $1$\\
1 & $1/\sqrt3$ & $1/4$ & --- & $1$\\
  & $-1/\sqrt3$ & $5/4$ & --- & $1$\\
\midrule
2 & $1$ & $0$ & --- & $3/4$\\
  & $-1/3$ & $2k/3$ & $0,1,2$ & $3/4$\\
\midrule
3 & $\pm1$ & $0$ & --- & $2/3$\\
  & $0$ & $k/2$ & $0,\ldots,3$ & $2/3$\\
\midrule
4 & $\pm1$ & $0$ & --- & $5/12$\\
  & $1/\sqrt5$ & $k/2$ & $0,\ldots,3$ & $25/48$\\
  & $-1/\sqrt5$ & $1/4+k/2$ & $0,\ldots,3$ & $25/48$\\
\midrule
5 & $\pm1$ & $0$ & --- & $1/2$\\
  & $1/\sqrt5$ & $2k/5$ & $0,\ldots,4$ & $1/2$\\
  & $-1/\sqrt5$ & $1/5+2k/5$ & $0,\ldots,4$ & $1/2$\\
\midrule
6 & $\pm1$ & $0$ & --- & $14/45$\\
  & $z_+$ & $k/2$ & $0,\ldots,3$ & $w_+$\\
  & $-z_+$ & $1/4+k/2$ & $0,\ldots,3$ & $w_+$\\
  & $-z_-$ & $k/2$ & $0,\ldots,3$ & $w_-$\\
  & $z_-$ & $1/4+k/2$ & $0,\ldots,3$ & $w_-$\\
\bottomrule
\end{tabular}
\end{table}

These effects satisfy the following completeness relations on each spin space\cite{LPT}:
\[
 \sum_x c_{q,x}^2
       \ket{q/2,n_{q,x}}\bra{q/2,n_{q,x}}=I_{q+1},
 \qquad \sum_x c_{q,x}^2=q+1.
\]
For $q=0$, the left-hand side consists of the single effect $1$. For $q=1$, the two directions are $\pm(1,1,1)/\sqrt3$,
and the corresponding states are orthogonal.
The $N$-copy effects in Eq.~\eqref{eq:finite-vidal-probability} follow by tensoring with $I_{\mu_{N,q}}$ and embedding in the corresponding Schur blocks~\cite{VidalMixed}.
\clearpage
\section{Implementation details of the measurement search}\label{app:search-details}
This appendix specifies the QR phase convention, componentwise Adam operations, and RCG line-search and termination rules used in Sec.~\ref{sec:barc}.

\subsection{QR phase adjustment and retraction properties}\label{app:search-qr}
To implement the QR retraction in Sec.~\ref{sec:barc-qr}, consider a thin QR factorization of a full-column-rank matrix $A\in\C^{s\times n}$:
\[
 \begin{aligned}
 A&=Q_0R_0,\qquad Q_0\in\C^{s\times n},\quad Q_0^{\dagger}Q_0=I_n,\\
 R_0&=(r_{ij})_{i,j=1}^n\in\C^{n\times n},\qquad r_{ij}=0\quad(i>j).
 \end{aligned}
\]
Full column rank implies $r_{ii}\ne0$. Define the diagonal unitary matrix $D\in\C^{n\times n}$ by
\[
 D:=\diag\left(\frac{r_{11}}{|r_{11}|},\ldots,\frac{r_{nn}}{|r_{nn}|}\right)
\]
and the phase-adjusted factors $Q\in\C^{s\times n}$ and $R\in\C^{n\times n}$ by
\[
 Q:=Q_0D,\qquad R:=D^{\dagger}R_0.
\]
They satisfy
\[
 A=QR,\qquad Q^{\dagger}Q=I_n,\qquad
 R_{ij}=0\ (i>j),\qquad R_{ii}=|r_{ii}|>0.
\]
This thin QR factorization with positive diagonal entries is unique,
and we define $\operatorname{qf}(A):=Q$.
For $T\in\mathrm T_Z\mathrm{St}_{\C}(s,n)$, the retraction is
\[
 \mathcal R_Z:\mathrm T_Z\mathrm{St}_{\C}(s,n)\longrightarrow\mathrm{St}_{\C}(s,n),\qquad
 \mathcal R_Z(T):=\operatorname{qf}(Z+T)
\]
as in Ref.~\cite{AMS}.
The tangent-space condition gives, for every $t\in\R$,
\[
 (Z+tT)^{\dagger}(Z+tT)=I_n+t^2T^{\dagger}T.
\]
In particular, $Z+tT$ has full column rank. Under the above phase convention,
$\mathcal R_Z(0)=Z$ and $\left.\frac{\mathrm d}{\mathrm dt}\mathcal R_Z(tT)\right|_{t=0}=T$.
Thus QR preserves the first-order update direction while restoring completeness to numerical precision.

\subsection{Adam conventions and default parameters}\label{app:search-adam}
For real arrays $A,B$ of the same shape and a component index $\alpha$, the componentwise operations are
\[
 (A^{\odot2})_\alpha:=A_\alpha^2,\qquad
 (A\oslash B)_\alpha:=\frac{A_\alpha}{B_\alpha}\quad(B_\alpha\ne0),\qquad
 (\sqrt B)_\alpha:=\sqrt{B_\alpha}\quad(B_\alpha\ge0).
\]
The expression $\sqrt{\widehat v_t}+\epsilon$ denotes addition of $\epsilon$ to every component.
We use $\beta_1=0.9$, $\beta_2=0.999$, and $\epsilon=10^{-8}$; the default learning rate is $\eta=0.005$, with $30000$ iterations.
The learning rate remains fixed during the Adam stage. The iterate after the prescribed number of updates initializes RCG.
The search terminates with a failure status if a nonfinite value occurs.
Iteration limits and learning rates for individual experiments are listed in Sec.~\ref{sec:experiment-conventions}.

\subsection{PR+ coefficient, line search, and stopping rules}\label{app:search-rcg}
With the notation of Sec.~\ref{sec:barc-rcg}, the PR+ coefficient, obtained by truncating the Polak--Ribi\`ere coefficient at zero, and the next search direction are
\begin{align}
 \beta_t^{\mathrm{PR+}}&=\max\left\{0,
 \frac{\langle g_{t+1},g_{t+1}-\mathcal T_tg_t\rangle}
 {\max(\langle g_t,g_t\rangle,10^{-30})}\right\},\label{eq:rcg}\\
 \Delta_{t+1}&=-g_{t+1}+\beta_t^{\mathrm{PR+}}\mathcal T_t\Delta_t.\nonumber
\end{align}
Whenever $\langle g_t,\Delta_t\rangle<0$ fails, reset the direction to $\Delta_t=-g_t$.
At each iteration, start with trial step size $\alpha=0.2$ and repeatedly halve it, accepting the first trial with a finite objective value that satisfies
\[
 f_W(\mathcal R_{Z_t}(\alpha\Delta_t))\le f_W(Z_t)+10^{-4}\alpha\langle g_t,\Delta_t\rangle.
\]
The accepted step is denoted by $\alpha_t$. The default limit is 40 trials.
Only the objective is evaluated at trial points; a new gradient is computed at the accepted point, avoiding gradient evaluations for rejected step sizes.
Convergence is declared when $\|g_t\|_{\mathrm F}<10^{-9}$, with a maximum of 2000 RCG iterations.
Termination at the iteration limit, line-search failure, and nonfinite values are distinguished from termination by the gradient criterion.
The designation PR+ does not imply exact Hessian conjugacy for general nonlinear problems.

\section{Proofs for the matrix-SOS hierarchy}\label{app:hierarchy-proofs}
We use the notation and support--kernel decomposition of Sec.~\ref{sec:effective-space}.

We use the Schur-complement criterion for positive definiteness.
For a Hermitian block matrix $H=\left(\begin{smallmatrix}H_{11}&H_{12}\\H_{12}^{\dagger}&H_{22}\end{smallmatrix}\right)$ with $H_{22}>0$, the Schur complement of $H_{22}$ is $H_{11}-H_{12}H_{22}^{-1}H_{12}^{\dagger}$, and
\[
 H>0\quad\Longleftrightarrow\quad
 H_{11}-H_{12}H_{22}^{-1}H_{12}^{\dagger}>0
\]
holds~\cite{BoydVandenberghe}.

\subsection{Proof of Lemma~\ref{lem:graded-degree}}\label{app:hierarchy-degree-proof}
\begin{proof}
\textup{(i)}\quad Split the SOS factors into support and kernel components:
\[
 \ket{f_\nu}=\begin{pmatrix}\ket{p_\nu}\\\ket{q_\nu}\end{pmatrix},\qquad
 \ket{p_\nu}\in\C[\xi]^r,\quad \ket{q_\nu}\in\C[\xi]^\ell
\]
Taking traces of the support and kernel blocks in Eqs.~\eqref{eq:Q} and \eqref{eq:blocks} gives
\begin{subequations}\label{eq:factor-traces}
 \begin{align}
 \sum_\nu\|p_\nu(\xi)\|^2
 &=w^m\Bigl(\tr X_0+\|\xi\|^2\tr \rho_0-2\sum_{i=1}^d\xi_i\tr A_{0i}\Bigr),
 \label{eq:support-factor-trace}\\
 \sum_\nu\|q_\nu(\xi)\|^2&=w^m\tr X_2
 \label{eq:kernel-factor-trace}
 \end{align}
\end{subequations}
When $\ell=0$, the equation for the kernel is omitted.

In general, for a finite family of polynomial vectors of the same dimension, $h_\nu(\xi)=\sum_{\alpha\in\mathbb N_0^d}h_{\nu,\alpha}\xi^\alpha$, with at least one nonzero member, set
\[
 t:=\max_{\nu,\alpha:\,h_{\nu,\alpha}\ne0}|\alpha|,\qquad
 h_\nu^{[t]}(\xi):=\sum_{|\alpha|=t}h_{\nu,\alpha}\xi^\alpha
\]
The homogeneous part of degree $2t$ in the sum of squared norms is $\sum_\nu\|h_\nu^{[t]}\|^2\not\equiv0$: its terms are nonnegative on the real domain, and at least one is nonzero. Thus $\deg\sum_\nu\|h_\nu\|^2=2t$.
Applying this observation to Eqs.~\eqref{eq:support-factor-trace} and \eqref{eq:kernel-factor-trace} yields $\deg p_\nu\le m+1$ and $\deg q_\nu\le m$. The case in which all vectors vanish is immediate.
Hence $\ket{f_\nu}=V_mc_\nu$ with $c_\nu\in\C^{N_m}$, and the Gram representation~\eqref{eq:matrix-sos-gram} proves (i).

\Needspace{10\baselineskip}
\medskip
\noindent\textup{(ii)}\quad Set $E:=\begin{pmatrix}I_r\\0\end{pmatrix}$, $K_i:=(A_{0i}\ \ 2A_{1i})$, and
$\widetilde V_0:=(I_n,\xi_1E,\ldots,\xi_dE)$.
\par\noindent For sufficiently large~$\lambda$,
\[
 \widetilde G_0:=\begin{pmatrix}
 \lambda I_n&-K_1^{\dagger}&\cdots&-K_d^{\dagger}\\
 -K_1&\rho_0&&0\\
 \vdots&&\ddots&\\
 -K_d&0&&\rho_0
 \end{pmatrix},\qquad
 \lambda I_n-\sum_{i=1}^dK_i^{\dagger}\rho_0^{-1}K_i>0
\]
so the Schur-complement criterion gives $\widetilde G_0>0$.
Since $EK_i+K_i^{\dagger}E^{\dagger}=2A_i$, we have
$\widetilde V_0\widetilde G_0\widetilde V_0^{\dagger}=Q_{B,\lambda I_n}$.

Expand
\[
 w^m=\sum_{|\alpha|\le m}c_\alpha\xi^{2\alpha},\qquad
 c_\alpha:=\frac{m!}{(m-|\alpha|)!\prod_{i=1}^d\alpha_i!}>0
\]
and, for each $|\alpha|\le m$, define the coordinate embedding $J_\alpha\in\R^{N_m\times(n+dr)}$ by
$V_mJ_\alpha=\xi^\alpha\widetilde V_0$.
The columns of $\xi^\alpha\widetilde V_0$ over all $|\alpha|\le m$ contain every monomial vector of degree at most $m$ in all components and every monomial vector of degree $m+1$ in the support components. Hence
$\sum_{|\alpha|\le m}\ran J_\alpha=\C^{N_m}$. Define
\[
 G_m:=\sum_{|\alpha|\le m}c_\alpha J_\alpha\widetilde G_0J_\alpha^{\dagger}\ge0,
 \qquad V_mG_mV_m^{\dagger}=w^mQ_{B,\lambda I_n}
\]
Since $c_\alpha>0$ and $\widetilde G_0>0$,
\[
 \ker G_m
 =\bigcap_{|\alpha|\le m}\ker J_\alpha^{\dagger}
 =\left(\sum_{|\alpha|\le m}\ran J_\alpha\right)^\perp
 =\{0\}.
\]
Therefore $G_m>0$, and $X^{(m)}:=\lambda I_n$ gives the required feasible point.
\end{proof}

\subsection{Proof of Lemma~\ref{lem:matrix-denominator}}\label{app:matrix-denominator-proof}
\begin{proof}
We use the SOS theorem with a denominator for real symmetric matrix polynomials of Dinh, Ho, and Le~\cite{DHL}. For the complex Hermitian case, consider the realification
\[
 \mathcal R(F):=\begin{pmatrix}
 \operatorname{Re}F&-\operatorname{Im}F\\
 \operatorname{Im}F&\operatorname{Re}F
 \end{pmatrix},\qquad
 S:=\frac1{\sqrt2}\begin{pmatrix}I_p\\-iI_p\end{pmatrix},\qquad
 S^{\dagger}\mathcal R(F)S=F
\]
Since $F(z)>0$ implies $\mathcal R(F)(z)>0$, apply the real symmetric result to $\mathcal R(F)$ and multiply the resulting SOS identity on the left by $S^{\dagger}$ and on the right by $S$.
\end{proof}

\subsection{Proof of Theorem~\ref{thm:denom-convergence}}\label{app:hierarchy-convergence-proof}
\begin{proof}
Take an optimal solution $X_*$ from Theorem~\ref{thm:exact-dual}, and set $X_\delta=X_*+\delta I_n$ with $\delta>0$.
Write their blocks as $X_{a,*}$ and $X_{a,\delta}$, respectively, for $a=0,1,2$.
For $t\in\R$ and $y\in\R^d$, set $z=(t,y)\in\R^{d+1}$ and $D_t=\diag(tI_r,I_\ell)$, and define
\begin{equation}\label{eq:compactification}
 \HH_X(t,y)=\begin{pmatrix}
 t^2X_0+\|y\|^2\rho_0-2tA_0(y)&tX_1-2A_1(y)\\
 tX_1^{\dagger}-2A_1(y)^{\dagger}&X_2
 \end{pmatrix}
\end{equation}
For $t\ne0$, $\HH_X(t,y)=D_tQ_X(y/t)D_t$.
By continuity, $\HH_{X_*}\ge0$ also holds at $t=0$.
Moreover, $\HH_{X_\delta}=\HH_{X_*}+\delta\diag(t^2I_r,I_\ell)$,
which is positive definite for $t\ne0$.
For $t=0$ and $y\ne0$, write $\ket{\psi}=\ket{a}\oplus\ket{b}$ with support and kernel components $\ket{a}\in\C^r$ and $\ket{b}\in\C^\ell$. If its quadratic form vanishes, then
\[
 0=\bra{\psi}\HH_{X_\delta}(0,y)\ket{\psi}
 =\bra{\psi}\HH_{X_*}(0,y)\ket{\psi}+\delta\|b\|^2
 \ge\delta\|b\|^2
\]
implies $\ket{b}=0$. It follows that $0=\|y\|^2\bra{a}\rho_0\ket{a}$, and $y\ne0$ and $\rho_0>0$ imply $\ket{a}=0$.
Hence $\HH_{X_\delta}(z)>0$ for every $z\ne0$.

When $\ell>0$, the support, cross, and kernel blocks of $\HH_{X_\delta}$ are homogeneous of degrees $2,1,0$, respectively, so the matrix as a whole is not homogeneous.
To obtain the homogeneity required by Lemma~\ref{lem:matrix-denominator}, set $X_{2,\delta}=X_{2,*}+\delta I_\ell>0$ and $L(t,y)=tX_{1,*}-2A_1(y)$, and take the Schur complement
\[
 P_\delta(t,y)=t^2X_{0,\delta}+\|y\|^2\rho_0-2tA_0(y)-L(t,y)X_{2,\delta}^{-1}L(t,y)^{\dagger}
\]
This is a \emph{homogeneous quadratic} matrix polynomial that is positive definite for $z\ne0$.
Apply Lemma~\ref{lem:matrix-denominator} to $P_\delta$ and set $t=1$. For some $m$, the polynomial $w(\xi)^mP_\delta(1,\xi)$ is then a matrix SOS.
In the identity obtained by completing the square,
\begin{equation}\label{eq:schur-sos-completion}
 Q_{B,X_\delta}(\xi)=\begin{pmatrix}P_\delta(1,\xi)&0\\0&0\end{pmatrix}
 +\begin{pmatrix}L(1,\xi)X_{2,\delta}^{-1/2}\\X_{2,\delta}^{1/2}\end{pmatrix}
  \begin{pmatrix}L(1,\xi)X_{2,\delta}^{-1/2}\\X_{2,\delta}^{1/2}\end{pmatrix}^{\dagger}
\end{equation}
the second term is also a matrix SOS. Since $w^m$ is a scalar SOS, it follows that $w^mQ_{B,X_\delta}$ is a matrix SOS.
If $\ell=0$, apply the same lemma with $P_\delta:=\HH_{X_\delta}$.
Proposition~\ref{prop:hierarchy-monotonicity} therefore gives $h_B\le d_m(B)\le\tr X_\delta=h_B+n\delta$ for all sufficiently large $m$. Letting $\delta\downarrow0$ proves the result.
\end{proof}
\begingroup
\small
\raggedright

\endgroup
\end{document}